\documentclass[%
 preprint, 
 amsmath,amssymb,
 aps, physrev,
superscriptaddress]{revtex4-2}
\usepackage{physics}
\usepackage{extarrows}

\usepackage{graphicx}
\usepackage{dcolumn}
\usepackage{bm}
\usepackage{tikz}
\usepackage{amsmath}
\usepackage{comment}

\usepackage{enumitem}

\newcommand{\newphi}{\phi}

\newcommand{\mo}{\mathcal{O}}

\usepackage{hyperref}

\usepackage{lipsum}
\usepackage{xcolor}
\usepackage{mathrsfs}

\begin{document}


\title{Acceleration radiation and HBAR thermodynamics for atoms falling into a
BTZ black hole: A CQM quantum-optics approach}

\author{H. E. Camblong}
\affiliation{Department of Physics and Astronomy,
University of San Francisco,
San Francisco, California 94117-1080, USA}

\author{A. Chakraborty}
\affiliation{Institute for Quantum Computing,
University of Waterloo,
Waterloo, ON, N2L 3G1, Canada}

\author{C. R. Ord\'{o}\~{n}ez}
\email{cordonez@central.uh.edu}
\affiliation{Department of Physics,
University of Houston,
Houston, Texas 77204-5005, USA}

\author{M. O. Scully}
\affiliation{The Institute for Quantum Science and Engineering,
Texas A\&M University,
College Station, TX 77843, USA}

\author{G. Valdivia-Mera}
\affiliation{Department of Physics,
University of Houston,
Houston, Texas 77204-5005, USA}
\affiliation{Texas Center for Superconductivity,
University of Houston,
Houston, TX 77204, USA}

\author{H. Wang}
\affiliation{The Institute for Quantum Science and Engineering,
Texas A\&M University,
College Station, TX 77843, USA}


\date{\today}\vspace{5mm}
\begin{abstract}
Atoms falling freely into a Ba\~{n}ados-Teitelboim-Zanelli (BTZ) black hole 
in a Boulware-like vacuum 
 are shown to emit radiation with a Planck spectrum at the Hawking temperature $T_{H}$. 
 This leads to thermal Hawking-like radiation for a cloud of falling atoms prepared 
 with random initial times. Moreover, the radiation is related to the relative equivalence principle, with the vacuum field modes accelerated with respect to the falling atom. 
The physics of the atom-field interactions is most easily described within a quantum optics approach, where each atom can be interpreted as a detector.
 Despite the topological nature of gravity in $(2+1)$ dimensions, 
 the thermodynamic and radiation properties of BTZ black holes are 
 still universally governed by the same near-horizon conformal quantum mechanics (CQM) applicable to higher-dimensional gravity. 
 This universal conformal behavior is exhibited by all fields in the background 
 of generic black holes, and generates an HBAR entropy $S_{\mathcal P}$ 
 associated with the photon radiation field that mimics the Bekenstein-Hawking entropy
 $S_{\mathrm{BH}}=A/4$, proportional to the black-hole horizon area, and with 
 the correct $1/4$ proportionality factor.
\end{abstract}

\maketitle
\newpage

\section{Introduction: Black hole thermodynamics and acceleration radiation}
\label{sec:introduction}

In the past few decades, research into the nature of black holes has revealed 
a universal set of relations linking aspects of gravitation, quantum physics, 
thermodynamics, and information theory in the presence of horizons as boundaries limiting causal 
access~\cite{hawking76, birrell-davies, BHT-review_Wald-2001, BHT-review_Padmanabhan-2010, BHT-review_Carlip-2014, BHT-review_Witten-2024}.

This framework, known as black hole thermodynamics, is
 based on the three pillars:
the Bekenstein-Hawking entropy of a black 
hole~\cite{bekenstein-S_1972, bekenstein-S_1973},
   \begin{equation}
  S_{\mathrm{BH}}
   = \frac{1}{4} 
   \frac{ k_{B}c^3}{ \hbar G }
\, A
    \; , 
    \label{eq:BH-entropy}
  \end{equation} 
proportional to its event horizon area $A$;
the area theorem and four geometric laws of black hole 
mechanics~\cite{BH-area_Hawking-1971, BH4_Bardeen-Carter-Hawking-1973}, 
along with the generalized second law of thermodynamics 
(GSL)~\cite{bekenstein-S_1973, bekenstein-S_1974};
and quantum black hole radiance~\cite{hawking74,hawking75} 
 with the Hawking temperature
\begin{equation}
T_H =
\frac{1}{2 \pi} \frac{\hbar }{ k_{B} c } \kappa
\;  ,
\label{eq:Hawking-temperature}
\end{equation}
 which is proportional to the surface gravity $\kappa$~\cite{GR_Carroll-2003}.
  In Eqs.~(\ref{eq:BH-entropy})--(\ref{eq:Hawking-temperature}),
    the Planck constant $\hbar$, speed of light $c$, and Boltzmann constant $k_{B}$ highlight    
  the quantum-relativistic and thermal aspects of these nontrivial spacetime effects.
Most importantly, the apparent universality of the connections involved in 
black hole thermodynamics suggests deeper insights
into the nature of spacetime and possible pointers towards a theory of
quantum gravity~\cite{QGravity-review}.

These results are
valid for a generic spacetime dimensionality $D$, as is well-known for $D \geq 4$.
But the case of lower dimensional gravity ($D <4$) calls for additional examination
and is the main focus of our paper. Here, we will consider
an alternative way to probe the thermodynamic properties of black holes via acceleration radiation, using the physics of the Unruh effect~\cite{Fulling-1973_CFQ,Davies-1975_Unruh,unruh-notes,ufd}.
Specifically, a thought experiment involving acceleration radiation from atoms falling into a black hole in an analog quantum-optics system 
provides an insightful probe 
of these phenomena for $D=3$.

\subsection{Acceleration radiation}

The Unruh effect~\cite{Fulling-1973_CFQ,Davies-1975_Unruh,unruh-notes,ufd}
involves analogous concepts for accelerated systems, including
the emission of acceleration radiation with the Unruh temperature
 \begin{equation}
T_U 
= \frac{1}{2 \pi} \frac{\hbar }{ k_{B} c } a
\;  ,
\label{eq:Unruh-temperature}
 \end{equation}
     proportional to the acceleration $a$.

These phenomena are naturally connected through the equivalence principle. In the model introduced in 2018 by Scully and collaborators~\cite{Scully_2018_HBAR}, depicted in Fig.~1, atoms are randomly injected and allowed to fall freely toward the event horizon of a four-dimensional Schwarzschild black hole, with boundary conditions that simulate a Boulware-like vacuum~\cite{birrell-davies,BV_Boulware-1975} adapted to stationary Schwarzschild coordinates. Although formulated in the full four-dimensional spacetime, the original model considered radial infall, thereby isolating an effective near-horizon $(t,r)$ sector that captures the essential acceleration-radiation dynamics~\cite{Scully_2018_HBAR}. This effective $(1+1)$-dimensional structure was first made systematic and explicitly calculational within the near-horizon/CQM framework of Ref.~\cite{acceler-rad-Schwarzschild}, where expansions of both the field modes and the geodesic trajectories were inserted directly into the atomic transition amplitudes. Although each freely falling atom occupies a locally inertial frame, it is accelerated relative to the Boulware field modes and therefore emits acceleration radiation with a spectrum characterized by the Hawking temperature~(\ref{eq:Hawking-temperature}).
%
  \begin{figure}[h]
    \centering
    \includegraphics[width=0.75\linewidth]{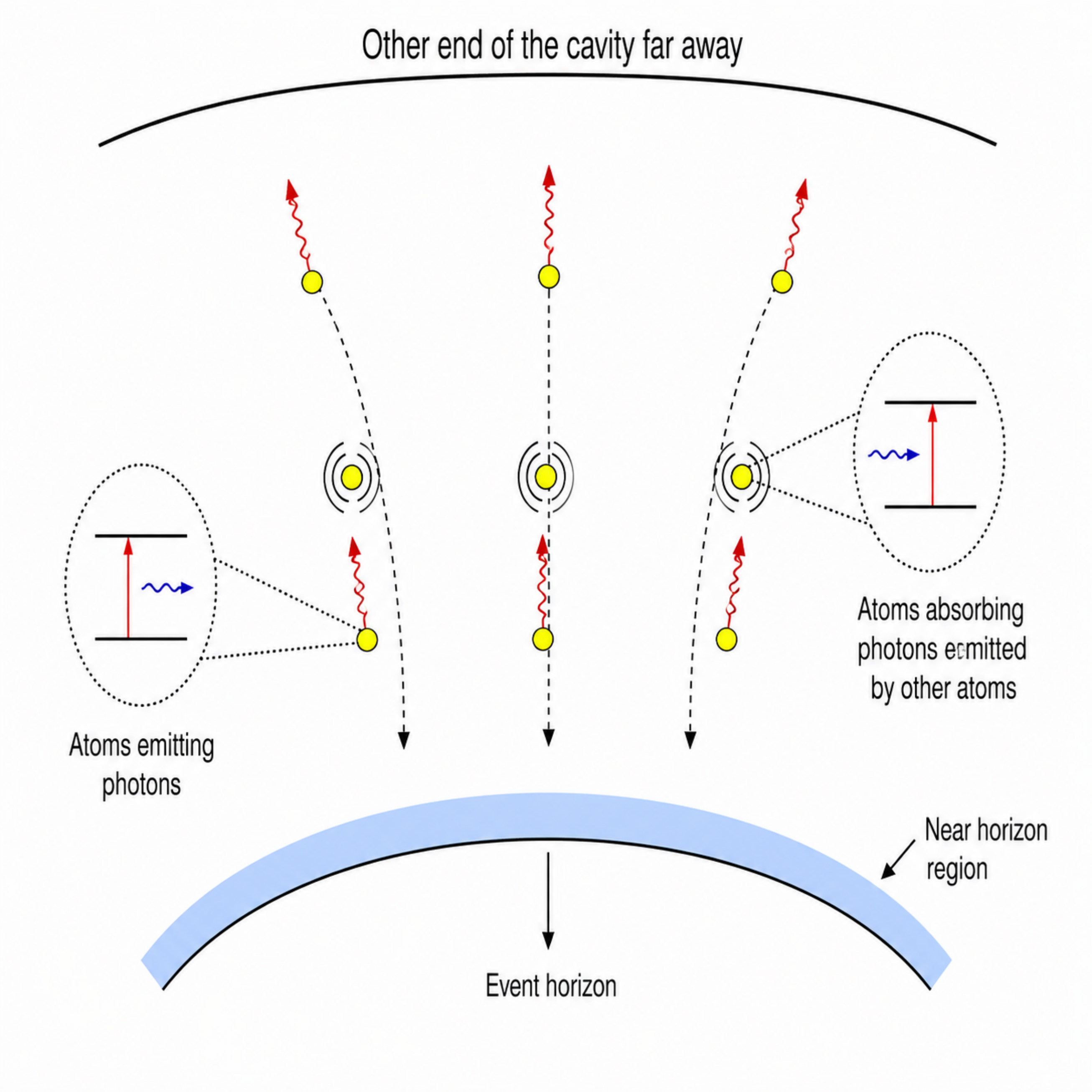}
    \caption{In the HBAR model, atoms follow free-fall paths 
    into a black hole.
    The state of the field, a Boulware vacuum, 
    simulates an analog quantum-optics system with boundary mirrors.}
       \label{fig:HBAR-setup}
\end{figure}
The resulting horizon-brightened acceleration radiation (HBAR)
provides a link between the Hawking and Unruh effects
with identical temperatures~(\ref{eq:Hawking-temperature}) 
and (\ref{eq:Unruh-temperature}):
 \begin{equation}
T_{\text{HBAR}}  =
 T_U \bigr|_{a = \kappa} 
= T_{H}
\;  .
\label{eq:HBAR-temperature}
 \end{equation}
A comparison between HBAR and acceleration radiation in inertial frames was subsequently carried out in Ref.~\cite{Ben-Benjamin-etal_2019_Unruh-rev}.
As shown in Refs.~\cite{Scully_2018_HBAR, Ben-Benjamin-etal_2019_Unruh-rev},
this problem can be easily tackled with the generic approach
of~\cite{AR_Scully-2003, AR_Belyanin-2006},
where quantum-optics techniques were adapted to spacetime field theory problems to understand the various configurations of accelerated two-state systems as detectors.  
This approach is based on the standard use of field quantization in curved spacetime~\cite{birrell-davies}, 
where the fields interact with two-level atoms
(Fig.~\ref{fig:HBAR-setup}) and settle into a steady-state configuration 
with apparent thermodynamic behavior~\cite{Scully_2018_HBAR}. 
Specifically, the steady state is established when
the emission and absorption processes satisfies the detailed-balance condition, with the ratio of their
probabilities given by the Boltzmann factor
\begin{equation}
\frac{P_{e, {\boldsymbol{s}} }   }{ P_{a, {\boldsymbol{s}} }  }
 = e^{-\beta_{H}\hbar \tilde{\omega}}
 \; 
 \label{eq:ratio_em_ab_Boltzmann}
 \end{equation}
where $\beta_H=(k_B T_H)^{-1}$ is the inverse Hawking temperature.
(In general, this method generates the corotating frequency
$ \tilde{\omega}$ associated with the black hole's angular velocity $\Omega_{H}$.
See Eq.~(\ref{eq:BH_ang-velocity}) and
as discussed in the main body of this work.)
Furthermore, the reduced field 
density matrix reveals the full-fledged character of
the thermodynamic-state condition.
This analysis---which we further develop in this article for the 
 geometry of black holes in $(2+1)$-dimensional spacetime---implies the emergence of a 
complete set of relations of HBAR thermodynamics, which 
can be shown to be in one-to-one correspondence with ordinary black hole 
thermodynamics~\cite{acceler-rad-Qopt-1, acceler-rad-Qopt-2}.
For example, the
HBAR entropy ${S}_{\mathcal P}$ of the radiation field changes according to~\cite{Scully_2018_HBAR}
\begin{equation}
    \dot{S}_{\mathcal P}
        = \frac{1}{4} 
   \frac{ k_{B}c^3}{ \hbar G }
      \bigl| \dot{A}_{\mathcal P} \bigr|
     \label{eq:HBAR_final}
     \; ,
\end{equation}
in which
 $\bigl| \dot{A}_{\mathcal P} \bigr|$ is the corresponding absolute value of the
change in the event-horizon area ${A}_{\mathcal P} $ due to the emission of acceleration radiation.
 This indeed mirrors qualitatively and quantitatively the Bekenstein-Hawking 
 entropy~(\ref{eq:BH-entropy}), including the critical prefactor $1/4$.

\subsection{Near-horizon conformal quantum mechanics and dimensionalities}

Another insight into HBAR thermodynamics is that it is governed by
near-horizon conformal invariance, just as ordinary black hole thermodynamics.
 For the latter, conformal symmetries 
 have been shown to provide a statistical foundation for the Bekenstein-Hawking 
 entropy~\cite{BHS_Carlip01-1995, BHS_Carlip02-1997, strominger_nh-BHS, BHS_Carlip03-1999, BHS_Carlip04-1999}. 
  An approach that explicitly displays the emergence of black hole thermodynamics is based on
 conformal quantum mechanics (CQM), which is a $(0+1)$-dimensional conformal field 
 theory~\cite{dAFF,CQM-renormalization-PLA}. 
 In this approach, CQM describes the near-horizon approximation to the field modes~\cite{Padmanabhan_BH-ISP_01, Padmanabhan_BH-ISP_02, guptasen, nhcamblong, nhcamblong-sc, vaidya, CQM_Moretti-2002}.
The conformal symmetry experienced by all fields in the black hole background is expected from the physical near-horizon blueshift~\cite{parkerprl}. 
Most importantly, the Bekenstein-Hawking entropy is interpreted via a brick-wall 
model~\cite{QBH_thooft-1985} in terms of a singular statistical mode counting that requires 
renormalization~\cite{nhcamblong, nhcamblong-sc, nhcamblong-heat-kernel, nhcamblong-conformal-tightness}. 
Subsequently, with the full power of CQM, HBAR thermodynamics was similarly shown to apply
 to a variety of Schwarzschild-like metrics in $D \geq 4$ 
spacetime dimensions~\cite{acceler-rad-Schwarzschild}---not surprisingly, the basic physics 
can be naturally extended from $D=4$ spacetime dimensions 
to higher-dimensional spacetimes~\cite{BH-dim_Myers-Perry-1986, BH-dim_Emparan-Reall-2008}.
In addition, CQM can also be displayed via corotating coordinates 
for black holes with angular momentum, described by the Kerr metric~\cite{acceler-rad-Kerr}. 
Remarkably, these findings confirm the universality of near-horizon CQM as the governing dynamics 
of black hole and HBAR thermodynamics for generic spacetime dimensionalities $D \geq 4$. 
These issues are examined in the centennial article of Ref.~\cite{QM-HBAR_centennial},
stressing the effectiveness of quantum physics applied to aspects of spacetime.
However, at first sight, the existing derivations of HBAR and near-horizon
CQM are not obviously transferable to lower spacetime dimensionalities ($D <4$).
This is the central issue that we are addressing in this paper.

\subsection{Lower-dimensional gravity and BTZ spacetime}

In this article, we extend the HBAR model and its associated thermodynamics to lower spacetime dimensionalities
 by focusing on the celebrated $(2+1)$-dimensional 
 Ba\~{n}ados-Teitelboim-Zanelli (BTZ) black hole~\cite{3DBH_BTZ-1992, 3DBH-geom_BTZ-1993}. 
 We thus show that the basic properties of HBAR physics work for the BTZ geometry in a manner 
 similar to the higher-dimensional black holes of Refs.~\cite{acceler-rad-Schwarzschild, acceler-rad-Kerr},
  leading to an emergent set of thermodynamic relations, and additionally highlighting their origin from near-horizon
 conformal quantum mechanics (CQM).
 The BTZ spacetime includes a rotating black hole in $(2+1)$-dimensional spacetime: 
  this is a black hole solution with mass $M$, angular momentum $J$, and
  a negative cosmological constant $\Lambda=-1/l^2$ (with curvature radius $l$).
 Its metric, in units with $c=1$, explicit Newton's gravitational constant $G$, and metric signature $(-,+,+)$,
 is given by the line element~\cite{3DBH_BTZ-1992, 3DBH-geom_BTZ-1993}.
\begin{equation}
 ds^2=-f(r)dt^2+f^{-1}(r)dr^2+r^2
    \bigl[ d\newphi - \varpi (r) dt \bigr]^2
    \; , 
    \label{eq:metric-BTZ}
\end{equation}
where the auxiliary functions
\begin{align}
   &
   f(r) \equiv (N^{\perp})^{2} (r)
   =-8 GM+\frac{r^2}{l^2}+\frac{16 G^2 J^2}{r^2}
\label{eq:f-factor}
   \\
   &
    \varpi (r) 
   \equiv - N^{\newphi}(r)= \frac{4 G J}{r^2}
   \; ,
 \label{eq:local_ang-velocity}
\end{align}
with $-\infty < t < \infty$, $0<r<\infty$, and $0 \leq \newphi < 2 \pi$,
correspond to the squared lapse function $(N^{\perp})^{2}$ 
and the angular shift $N^\newphi(r)$ of the 
Arnowitt–Deser–Misner (ADM) formalism~\cite{ADM_1959, ADM-repub_2008}.
The condition $f(r)=0$ defines the black hole horizons, with radial-coordinate roots $r_{\pm}$ 
(see Sec.~\ref{sec:BTZ-spacetime})~\cite{3DBH_BTZ-1992, 3DBH-geom_BTZ-1993,BTZ-BH_Carlip-1995}.
Equation~(\ref{eq:metric-BTZ}) can be interpreted in terms of 
 frame dragging in a local corotating frame with position-dependent angular velocity $\varpi$ 
relative to an external reference system~\cite{Landau-Lifshitz_CTFs, MTW-gravitation, frolov}. 
As the outer event horizon $r=r_{+}$ is approached, $\varpi$ becomes the angular velocity of the black hole,
\begin{equation}
\Omega_{H} =
\lim_{r \rightarrow r_{+}} \varpi
=
 \frac{4GJ}{r_{+}^2 }
 =
 \frac{r_{-}}{r_{+} l}
\; .
\label{eq:BH_ang-velocity}
\end{equation}

An important motivation for the analysis of the BTZ black hole is the expected simplicity of lower-dimensionality gravity ($D<4$), which makes it an ideal laboratory for foundational questions of gravitation and quantum
gravity~\cite{3DQGravity-Carlip-1998, 3DQGravity-Carlip-2005, 3DQGravity-Carlip-2024}.
In fact, inspection of Eqs.~(\ref{eq:metric-BTZ})--(\ref{eq:local_ang-velocity}) 
shows similarities with the known higher-dimensional Schwarzschild and Kerr 
spacetimes~\cite{GR_Carroll-2003, Landau-Lifshitz_CTFs, MTW-gravitation, frolov}, 
but with a reduced number of coordinates. This suggests a pathway to BTZ black hole thermodynamics,
 even though the $(2+1)$-dimensional solution does not have all the properties of higher-dimensional black holes.
 In particular, the black hole {\it only exists for negative cosmological constant\/}
 and it does not have a curvature singularity in the usual 
 sense, as all the curvature invariants remain finite~\cite{3DBH_BTZ-1992, 3DBH-geom_BTZ-1993}---instead,
 it still has a quasi-regular singularity with geodesic incompleteness, a conical singularity, and
a distributional (delta-function) singularity associated with Wilson loops~\cite{BTZ_Briceno-etal-2024}.
Most importantly, this formal analogy implies the existence of an event horizon and a near-horizon universal behavior 
that leads to the emergence of standard thermodynamic entropy and related thermodynamic 
properties~\cite{BHT-review_Padmanabhan-2010}.
These are horizon properties whose existence we will further explore in this article.

In this context, confirming a {\it fully consistent picture\/},
various aspects of {\it BTZ black hole thermodynamics\/} have been derived 
via an arsenal of techniques:
field-mode expansions~\cite{BTZ-Tmodes_Hyun-1994, BTZ-Tmodes-Eucli_Ichinose-1995},
 entropy tunneling~\cite{BTZ-Ttunn-S_Englert-Reznik-1994},
 the Euclidean path integral 
  framework~\cite{BTZ-thermo_Carlip-Teitelboim-1995, BTZ-Tmodes-Eucli_Ichinose-1995},
the microcanonical Brown-York framework~\cite{BH-thermo-AdS_Brown-Creighton-Mann-1994},
 Wald's Noether charges~\cite{BTZ-BH_Carlip-1995, 3D-th-topo_Carlip-etal-1995},
 detailed balance of gravitational perturbations~\cite{AdS3-perturb_Emparan-Sachs-1998},
the radiation tunneling formalism~\cite{BTZ-Ttunn_Medved-2002},
and the anomaly cancellation method~\cite{BTZ-Tanom_Jiang-2007, BTZ-Tanom_Li-Li-Ren-2010}.
In addition, the BTZ black hole entropy, in agreement with the general result 
of Eq.~(\ref{eq:BH-entropy}), is successfully 
derived by counting microscopic degrees of freedom of two-dimensional conformal field theory 
at the horizon~\cite{BHS_Carlip01-1995, BHS_Carlip02-1997} or at asymptotic infinity~\cite{strominger_nh-BHS};
with subsequent generalizations to higher dimensions~\cite{BHS_Carlip03-1999, BHS_Carlip04-1999}. 
These results typically rely on the Cardy formula~\cite{Cardy-form_Cardy-1986, CFT-Cardy-form_Blote-etal-1986}
of conformal field theory~\cite{BTZ-openq_Carlip-1998, 3D-CFT-BH_Carlip-2005, Ad3/CF2_Kraus-2008},
using the central charge in asymptotically AdS$_{3}$ 
spacetimes~\cite{Brown-Henneaux-1986}.
Many directions and extensions of these basic frameworks have appeared over the years,
but open questions still remain, especially in regards to the relations between the different methods, the boundary conditions, and the microscopic degrees of freedom~\cite{BTZ-openq_Carlip-1998, 3D-CFT-BH_Carlip-2005, 3DQGravity-Carlip-2024}.
The results of this article provide further evidence supporting a consistent picture 
of black hole thermodynamics within lower-dimensional gravity.

 \subsection{Organization of this article}

The structure of this article is as follows. 
In Sec.~\ref{sec:BTZ-spacetime}, 
we review the basic geometric properties of the BTZ black hole spacetime,
starting from the metric~(\ref{eq:metric-BTZ})--(\ref{eq:local_ang-velocity}), and including
a complete treatment of the geodesics.
In Sec.~\ref{sec:BTZ-spacetime_FT},
 the corresponding field theory equations relevant for the acceleration radiation are developed,
 including the basic quantization relations.
In Sec.~\ref{sec:CQM_KGeq-geodesics},
we derive the near-horizon limit of the BTZ field equations and geodesics, emphasizing their
universal features and showing the governing role played by conformal quantum mechanics (CQM).
Section~\ref{sec:HBAR-thermodynamics} 
derives the properties of the horizon-brightened acceleration radiation (HBAR) emergent from
near-horizon CQM, including the radiation probability amplitudes and the BTZ thermal 
characterization using both detailed balance and density matrix arguments;
moreover, the corresponding HBAR entropy and the whole set of HBAR thermodynamics relations
are highlighted.
We conclude this article in Sec.~\ref{sec:conclusions} with remarks on the universality of our
results and related open problems.
Finally, the appendices give additional results on three major topics:
the BTZ geometry, the atom-field interactions and density matrix, and the 
Boulware vacuum.

Regarding the unit conventions for the remainder of this article, we will typically use natural units 
in which $c=1$, $\hbar =1$, and $k_B=1$, in addition to $8G =1$, which is the most common choice
for the BTZ spacetime following
Refs.~\cite{3DBH_BTZ-1992, 3DBH-geom_BTZ-1993}.
Ordinary units can be recovered by the 
simultaneous replacements $M \rightarrow  8GM$ and $J \rightarrow  8GJ$.

\section{BTZ Spacetime: Geometry}
\label{sec:BTZ-spacetime}

In this section, we survey properties of the spacetime geometry associated with 
a rotating black hole in $D=2+1$ dimensions. 
This is a BTZ black hole, for which the metric line element is given by
Eqs.~(\ref{eq:metric-BTZ})--(\ref{eq:local_ang-velocity}). 

Some remarks on lower-dimensionality gravity ($D<4$) are in order,
for context, 
highlighting the relevance of this extension.
First,
the action for this theory, $(2+1)$-dimensional Einstein gravity, is
\begin{equation}
S = 
\int d^3x \sqrt{|g|}  \left[ \frac{1}{2\kappa_{\mathrm E}} \bigl( R - 2\Lambda \bigr) 
+ \mathcal{L}_{M} \right] 
+ B
\; ,
\label{eq:Einstein-gravity-action}
\end{equation}
where $R$ is the Ricci scalar, $\Lambda$ is the cosmological constant, 
$\mathcal{L}_{M}$ is the matter Lagrangian density, and $B$ is a boundary term typically required 
to cancel surface integrals in the variational procedure leading to proper action 
extrema~\cite{Regge-Teitelboim-1974};
the Einstein constant $\kappa_{\mathrm E} = 8 \pi G/c^4$ becomes $\kappa_{\mathrm E} = \pi$ in the chosen natural units with $8G=1$.
The corresponding Einstein field equations become
\begin{equation} 
R_{\mu\nu} - \frac{1}{2} \left( R - 2 \Lambda \right)
g_{\mu\nu}  = \kappa_{\mathrm E} T_{\mu\nu}
\; .
\label{eq:EFE}
\end{equation}
Second, despite being naturally simpler, lower-dimensionality gravity
 encounters additional technical features and challenges.
 Specifically, in $D= 2+1$ dimensions, the peculiar properties and topological
  nature of Einstein gravity were analyzed in the seminal works of Deser, Jackiw, ’t Hooft, and Witten~\cite{3Dgravity_Deser-J-tH-1984, 3Dgravity_Deser-J-1984, 3DGravity-CS_Achucarro-1986, 3Dcone_Deser-J-1988, 3Dgravity_tHooft-1988, 3Dgravity_Witten-1988, 3Dgravity-topo_Witten-1989}, and it was
found that it can be reformulated as a Chern-Simons theory with the Poincaré group as its gauge group~\cite{3DGravity-CS_Achucarro-1986, 3Dgravity_Witten-1988}.
Moreover, in this reduced dimensionality, the Weyl tensor vanishes identically;
thus, the distribution of matter, as given by $T_{\mu \nu}$, completely determines 
the curvature tensor algebraically. As a result, outside matter regions, with $T_{\mu \nu} =0 $, 
the Riemann tensor vanishes and spacetime is locally flat for $\Lambda =0$, or
locally de Sitter or anti-de Sitter for $\Lambda >0$ or $\Lambda <0$ respectively.
From this analysis and the local counting of independent components, it follows that 
$(2+1)$ Einstein gravity has no local dynamical degrees of freedom, no corresponding Newtonian analog limit, and no gravitational waves~\cite{3Dgravity-Barrow-1986, 3Dgravity-Cornish-1991, BTZ-BH_Carlip-1995, 3D-CFT-BH_Carlip-2005}.
Third, while the discovery of the BTZ black hole was unexpected,
 its unusual properties~\cite{BTZ-BH_Carlip-1995, 3D-CFT-BH_Carlip-2005}
have generated an extensive literature in subsequent years, as
it can be used as a toy model to understand black hole thermodynamics and quantum gravity
in a simplified, lower-dimensional 
setting~\cite{BTZ-Tmodes_Hyun-1994, BTZ-Tmodes-Eucli_Ichinose-1995, BTZ-Ttunn-S_Englert-Reznik-1994, BTZ-thermo_Carlip-Teitelboim-1995, BTZ-Tmodes-Eucli_Ichinose-1995, BH-thermo-AdS_Brown-Creighton-Mann-1994, BTZ-BH_Carlip-1995, 3D-th-topo_Carlip-etal-1995, AdS3-perturb_Emparan-Sachs-1998, BTZ-Ttunn_Medved-2002, BTZ-Tanom_Jiang-2007, BTZ-Tanom_Li-Li-Ren-2010, BHS_Carlip01-1995, BHS_Carlip02-1997, strominger_nh-BHS, BHS_Carlip03-1999, BHS_Carlip04-1999, Cardy-form_Cardy-1986, CFT-Cardy-form_Blote-etal-1986, BTZ-openq_Carlip-1998, 3D-CFT-BH_Carlip-2005, Ad3/CF2_Kraus-2008, BTZ-openq_Carlip-1998, 3D-CFT-BH_Carlip-2005, 3DQGravity-Carlip-2024},
 as mentioned in Sec.~\ref{sec:introduction}.
Finally, one of the leading reasons for the recurrent interest in the BTZ geometry
is its relation to several higher-dimensional black holes;
specifically, in the near-horizon limit, 
many black hole solutions derived from string theory reduce
 to the form BTZ$\, \times \, M$, where $M$ is a simple manifold.
Then, the entropy of these stringy black holes can be obtained from the BTZ black hole solution directly or via duality~\cite{strominger_nh-BHS, Balasubramanian_nh-BHgeom-98, Birmingham_nh-BH-string-98, DeAlwis_BH-DBI-98, Kaloper_BH-DBI-98, Lee-Myung_BTZ-5D-98, MaldaStrom_AdS3-98, Sachs_BHT-98}.

\subsection{BTZ geometry: Basic metric properties}
\label{sec:BTZ-geometry_basic}

Despite the locally trivial structure of $(2+1)$-dimensional Einstein gravity, its globally nontrivial properties 
generate the BTZ black hole solution of the field equations~(\ref{eq:EFE})
when $T_{\mu \nu} =0$ and $\Lambda =-1/l^2$.
The BTZ black hole can be characterized with the metric of 
Eqs.~(\ref{eq:metric-BTZ})--(\ref{eq:local_ang-velocity}),
describing a stationary and axisymmetric spacetime, written in 
generalized Schwarzschild-like coordinates with an additional angular momentum $J$.
The mass and angular momentum parameters $M$ and $J$ are the conserved Noether charges 
associated with asymptotic symmetries under time displacements and rotational
invariance, respectively~\cite{Noether_Lee-Wald-1990}. 
Now, in $(2+1)$-dimensional spacetime, $M$ (i.e., more explicitly, $GM$) is dimensionless, while $J$ 
(i.e., more explicitly, $GJ$) has dimensions of length. 
A negative cosmological constant $\Lambda = - 1/l^2$ yields a length scale $l$ and generates
a BTZ spacetime of constant curvature that is locally isometric to anti-de Sitter (AdS$_{3}$).
An alternative form of the metric is obtained by expanding Eq.~(\ref{eq:metric-BTZ}) explicitly; 
again, in units with $8G = 1$,
\begin{equation}
\begin{aligned}
   &  ds^2=
    g_{tt} dt^2  +  2  g_{t \newphi} dt d \newphi  + g_{\newphi \newphi} d \newphi^2  +  g_{rr} dr^2 
    \; ,
\\
 &
   g_{tt} = - \left[ f(r) - \frac{J^2}{4r^2} \right]
   =-  \left(-M+\frac{r^2}{l^2}\right)  
        \; , \; \; \;
     g_{t \newphi} = - \frac{J}{2}
       \; , \; \; \;
      g_{\newphi \newphi}= r^2
           \; ,
\\
 &
    g_{rr} =  \left[ f(r) \right]^{-1} = \left(-M+\frac{r^2}{l^2}+\frac{J^2}{4r^2}\right)^{-1}
    \; ,
        \end{aligned}
    \label{eq:metric-BTZ_exp}
\end{equation} 
which further stresses similarities to the $(3+1)$-dimensional Kerr 
metric~\cite{GR_Carroll-2003, Landau-Lifshitz_CTFs, MTW-gravitation, frolov}, 
in Boyer-Lindquist coordinates.

The stationary and axisymmetric nature of the BTZ geometry is completely characterized 
by time and rotation symmetries; furthermore, the geometry is not static when $J \neq 0$.
This can be recognized by direct inspection of Eq.~(\ref{eq:metric-BTZ_exp}):
 the metric itself is independent of $t$ and $\newphi$, but the line element has cross terms 
 $dt\,d\newphi$ that are not invariant under time reversal for $J \neq 0$.
  The corresponding Killing vectors,
    \begin{equation}
\boldsymbol{\xi}_{(t)} \equiv \partial_t  
\; \; , \; \; \; \; \;
\boldsymbol{\xi}_{(\newphi)} \equiv \partial_\newphi 
\; ,
\label{eq:Killing-vectors_stationary}
\end{equation}
are the infinitesimal generators of metric
invariance~\cite{GR_Carroll-2003, MTW-gravitation, frolov, GR_Wald-1984}
in the given coordinate choices, and they provide a subset of coordinate basis vectors with
$\xi_{(t)}^{\mu} = \delta^{\mu}_{t}$ and $\xi_{(\newphi)}^{\mu} = \delta^{\mu}_{\newphi}$.
In particular, the Killing vectors give a convenient symmetry-based representation of the part of the
metric associated with their invariances, in the form
\begin{equation} 
g_{ab} = \boldsymbol{\xi}_{(a)} \!  \cdot  \! \boldsymbol{\xi}_{(b)}
\label{eq:metric-Killing}
\end{equation} 
[where $a,b$ label the coordinates $(t,\phi)$ in the submanifold $r=$ constant, 
with $g_{ab}$ being the corresponding induced metric].
Moreover, comparison of Eqs.~(\ref{eq:metric-BTZ})--(\ref{eq:local_ang-velocity}) and (\ref{eq:metric-BTZ_exp})
shows that the frame-dragging angular velocity 
\begin{equation} 
\varpi = - \frac{g_{t \newphi} }{g_{\newphi \newphi}} 
=\frac{J}{2r^2}
\; 
\label{eq:local_ang-velocity_8G=1}
\end{equation}
[here,
Eq.~(\ref{eq:local_ang-velocity}) is written with $8G=1$]
can be represented in the same 
symmetry-based framework as 
\begin{equation} 
\varpi
= -  \frac{  \boldsymbol{\xi}_{(t)} \!  \cdot  \! \boldsymbol{\xi}_{(\newphi)} }{  
\boldsymbol{\xi}_{(\newphi)} \!  \cdot  \! \boldsymbol{\xi}_{(\newphi)} }
\; .
\label{eq:local_ang-velocity_geom}
\end{equation}

Now, given this explicit frame dragging,
 the coordinate change 
\begin{equation}
\tilde{t}=t  
\; , \; \; \; 
\tilde{\phi}=\phi-\Omega_{H} t
\; ,
\label{eq:rotating-coords_separation_1}
\end{equation}
defines a frame that
is corotating with the black hole at its angular velocity $\Omega_{H}$
given by Eq.~(\ref{eq:BH_ang-velocity}).
Then, the Killing vectors associated with the coordinates~(\ref{eq:rotating-coords_separation_1}) 
include a corotating timelike Killing vector 
   \begin{equation}
\boldsymbol{\xi}_{(\tilde{t})} \equiv \partial_{\tilde{t}} 
=  \boldsymbol{\xi}_{(t)} + \Omega_{H} \boldsymbol{\xi}_{(\phi)}
\label{eq:Killing-vectors_corot-time}
\; ,
\end{equation}
 and a corotating angular Killing vector 
   \begin{equation}
\boldsymbol{\xi}_{(\tilde{\newphi})} \equiv \partial_{\tilde{\newphi}} 
=  \boldsymbol{\xi}_{(\phi)}
\label{eq:Killing-vectors_corot-ang}
\; .
\end{equation}
Additional metric relations and their impact on the geometry are discussed 
in Appendix~\ref{app:BTZ-geometry-additional}.

\subsection{BTZ geometry: Structure}
\label{sec:BTZ-geometry_structure}

For the BTZ black hole, two sets of surfaces can be identified as critical boundaries, in a manner
similar to the $(3+1)$-dimensional Kerr spacetime. 
They include:
\begin{enumerate}[label=(\roman*)]
\item
 The horizons ${\mathcal H}^{\pm}$ (outer and inner, for $r=r_{\pm}$), which 
define the existence of a BTZ black hole, and are identified via the vanishing 
of the metric scale factor 
\begin{equation}
   g^{rr} =  f(r)=
    {\frac {(r^{2}-r_{+}^{2})(r^{2}-r_{-}^{2})}{l^{2}r^{2}}}
    \; ,
    \label{eq:f(r)_factor}
\end{equation}
with roots
\begin{equation}
    r_\pm^2=\frac{M l^2}{2}
    \qty[
    1 \pm \sqrt{1-\left( \frac{J}{M l} \right)^2}]
    \; .
    \label{eq:BH_horizons}
\end{equation}
\item
 The static (stationary) limit or ergosurface ${\mathcal S}$, 
which is an infinite redshift surface with respect to an asymptotic stationary observer, 
identified by 
$g_{tt} = \boldsymbol{\xi}_{(t)} \cdot \boldsymbol{\xi}_{(t)} = 0$, 
with a value
$r=r_{e}$ given by
\begin{equation}
  r_{e}  = \sqrt{M}l=(r_+^2+r_-^2)^{1/2}
\label{eq:static-r_{e}}
\; .
\end{equation}
\end{enumerate}
In particular, when the roots $r_{\pm}$ in Eq.~(\ref{eq:BH_horizons}) are real, they
 select null surfaces that are known to be horizons from general properties of
stationary spacetimes~\cite{GR_Carroll-2003, GR_Wald-1984};
for the identification of these critical boundaries, 
see Appendix~\ref{app:BTZ-geometry-additional}, 
where it is also shown that the horizon condition $ g^{rr}(r) = 0$ is equivalent to
$ g_{\tilde{t}\tilde{t}} = \boldsymbol{\xi}_{(\tilde{t})} \cdot   \boldsymbol{\xi}_{(\tilde{t})}    =0$.
    Now, the roots $r_{\pm}$ in Eq.~(\ref{eq:BH_horizons}) are real
 when the conditions $M > 0$ and $\abs{J}  \leq   Ml$ are satisfied. Correspondingly,
as it happens for similar higher-dimensional black holes with additional conserved
charges (other than the mass $M$), the solution has two horizons~\cite{GR_Carroll-2003}. 
The outer value $r_{+}$ is an event
horizon, while the inner horizon at $r=r_{-}$ is of the Cauchy type~\cite{BTZ-BH_Carlip-1995}.
The threshold equality $\abs{J} = M l$ defines an extremal black hole, with $r_+ = r_-$, and
the nonextremal black holes satisfy $\abs{J}  <   Ml$.
If the threshold were exceeded, with
 $\abs{J}  >   Ml$, a naked singularity would form~\cite{BTZ_Briceno-etal-2024}.
For the analysis of interest in most practical applications, including acceleration radiation (as in this review article),
the nonextremal geometry is considered, for which
\begin{equation}
f'_{+} \equiv  f'(r_{+}) = \frac{2 (r_{+}^2-r_{-}^2) }{ r_{+}l^2 } 
=
\frac{2r_+}{l^2}-\frac{J^2}{2r_+^3}
\neq 0
\; ,
\label{eq:f-prime}
\end{equation}
 where the prime stands for the radial derivative.
Furthermore, from the algebraic invariants of the polynomial equation $f(r)=0$
  [Eqs.~(\ref{eq:metric-BTZ}) and (\ref{eq:metric-BTZ_exp})], 
  the conserved charges can be rewritten in terms of the horizon parameters $r_\pm$ and cosmological scale $l$ through 
\begin{equation}
    M=\frac{r_+^2+r_-^2}{l^2}\quad,\quad J=\frac{2r_+r_-}{l}
    \; .
    \label{eq:M-J_from_r-pm}
\end{equation}

In short, in the explicit form of Eq.~(\ref{eq:metric-BTZ_exp}),
there is a clear distinction between the time and radial parts of the metric: 
$ -g_{tt} \neq \left( g_{rr} \right)^{-1} = g^{rr}$.
(See Appendix~\ref{app:BTZ-geometry-additional}.)
The condition $g^{rr} (r) =0$ specifies the radial coordinate values~(\ref{eq:BH_horizons}) of the horizons, 
while the condition $g_{tt}(r) =0$ gives Eq.~(\ref{eq:static-r_{e}}) as the location $r=r_{e}$ of the 
ergosurface with an infinite redshift with respect to the time $t$ of a distant observer.
By the hierarchy of distances: $r_{-}<r_{+} < r_{e}$, there exists a region $ r_{+} < r < r_{e}$,
where particles cannot be at rest with respect to distant observers: the ergosphere---just 
as for the case of Kerr black holes~\cite{GR_Carroll-2003, MTW-gravitation, frolov, GR_Wald-1984}.
This behavior arises 
from the metric element $g_{tt} = \boldsymbol{\xi}_{(t)} \cdot \boldsymbol{\xi}_{(t)} $
(with time $t$ suitable for distant observers)
changing sign upon crossing the ergosurface, from $g_{tt} <0$ for $r>r_{e}$
to $g_{tt} >0$ for $r<r_{e}$.
In other words, the Killing vector $\boldsymbol{\xi}_{(t)}$
  turns from timelike to spacelike when entering the ergosphere, where it cannot be used
  to describe time evolution; instead, time evolution inside the ergosphere, approaching the event horizon,
 is governed by the corotating timelike Killing vector $\boldsymbol{\xi}_{(\tilde{t})}$.
 Moreover, this also has implications for the field theory described in Sec.~\ref{sec:FT-BTZ-geom},
for which it is the Killing vector
 $\boldsymbol{\xi}_{(\tilde{t})}$
that provides the correct differential operator with respect to which positive and negative frequency modes 
of the field are selected to form a complete orthonormal basis 
of the field operator.
 The analysis with geodesics in the next section further elaborates on the properties of the ergosphere.

\subsection{BTZ geometry: Additional black hole properties and geodesics}
\label{sec:BTZ-geometry_additional}

  There are two relevant geometrical quantities that play a central role in black hole thermodynamics
and in HBAR thermodynamics, and the BTZ black hole is no exception. 
The first one is the event horizon area 
\begin{equation}
   A = \int \sqrt{g_{\newphi\newphi}(r_+)} \, d\newphi = 2\pi r_+
   \; ,
 \label{eq:EH-area}
\end{equation}
as follows from Eq.~(\ref{eq:metric-BTZ_exp}).
This quantity determines the Bekenstein-Hawking entropy $S_{\mathrm{BH}}$ in Eq.~(\ref{eq:BH-entropy}),
 where the ``area'' for two spatial dimensions reduces to the circumference $ 2\pi r_+$.
The second one is the surface gravity $\kappa$, 
defined with the covariant derivatives of the timelike Killing vector 
 $\boldsymbol{\xi} \equiv \boldsymbol{\xi}_{(\tilde{t})} $ in Eq.~(\ref{eq:Killing-vectors_corot-time}), 
 which is normalized to conform to an operationally defined notion of gravitational 
 acceleration~\cite{GR_Carroll-2003, frolov}
measured from infinity with $\boldsymbol{\xi}_{({t})} $.
Specifically, from the general definition
and a straightforward calculation, the BTZ surface gravity is
\begin{equation}
\kappa = 
\sqrt{
\left.
-\frac{1}{2} \left( \nabla_{\mu} {\xi}_{\nu} \right) 
\left(  \nabla^{\mu} {\xi}^{\nu} \right)
\right|_{r=r_{+}}
\! \!
}
\; 
  = \frac{f'_{+}}{2} =
   \frac{r_+^2 - r_-^2}{l^2 r_+}
\; .
\label{eq:surface-gravity_BTZ}
\end{equation}
Moreover, $\kappa$
determines the Hawking temperature $T_{H}$ in Eq.~(\ref{eq:Hawking-temperature}).

 Finally, the following discussion of the BTZ geodesics provides the ingredients needed 
 for a description of the free-fall spacetime trajectories of atoms in a background metric, 
 leading to the acceleration radiation in Sec.~\ref{sec:CQM_KGeq-geodesics}.
In addition, these results constrain the kinematic behavior of particles in the ergosphere, 
including their motions in the all-important near-horizon region.
The geodesics of a particle are governed by second-order equations, but these can be integrated to first-order forms
using the constraints or constants of the motion associated with a particle's spacetime velocity 
$\boldsymbol{U}$ as the 
geodesic tangent vector~\cite{GR_Carroll-2003, MTW-gravitation, frolov, GR_Wald-1984}, with
components $U^\nu = dx^\nu/d\tau$. This is proportional to the particle's spacetime 
momentum $\boldsymbol{p}$, and they are the same when considered per unit mass.
Then, for the BTZ geometry, these constraints are provided by: 
(i) the two conserved quantities
\begin{align}  
    &
    e = -   \boldsymbol{\xi}_{(t)} \cdot \boldsymbol{U} = -U_{t}  
 =
     -g_{tt} \frac{dt}{d\tau} 
- g_{t\newphi} \frac{d\newphi}{d\tau} 
\; ,
      \label{eq:Killing-energy} 
      \\
      &
      \ell
  = \boldsymbol{\xi}_{(\phi)} \cdot \boldsymbol{U}
   = U_{\phi}
       = g_{\newphi t} \frac{dt}{d\tau} 
+ g_{\newphi\newphi} \frac{d\newphi}{d\tau}
    \; ,
    \label{eq:Killing-ang-mom} 
\end{align}
corresponding to the Killing vectors~(\ref{eq:Killing-vectors_stationary}),
i.e., the dimensionless energy per unit mass $e$ and the angular momentum per unit mass $\ell$ 
(with units of length) respectively; 
and (ii) the velocity normalization
 \begin{align}
\boldsymbol{U} \cdot \boldsymbol{U}
=
g_{rr} ( U^{r})^2 + g^{ab} U_{a} U_{b} 
= g_{rr} \qty(  \frac{dr}{d\tau} )^2 + 
g^{tt} e^2 + g^{\newphi \newphi} \ell^2 - 2 g^{t \newphi} e \ell
= -1 
\label{eq:velocity-normalization}
 \end{align}
[from Eq.~(\ref{eq:metric-Killing}), with $a,b$ from $(t,\phi)$],
 along with Eqs.~(\ref{eq:Killing-energy})--(\ref{eq:Killing-ang-mom})].
Moreover, 
the metric element $g_{tt} = \boldsymbol{\xi}_{(t)} \cdot \boldsymbol{\xi}_{(t)} $ changes sign upon crossing the 
ergosurface, from $g_{tt} <0$ for $r>r_{e}$ to 
 $g_{tt} >0$ for $r<r_{e}$, thus having a time Killing vector that becomes of positive norm inside the ergosphere.

The geodesic first-order equations for $dt/d\tau$ and $d\newphi/d\tau$ 
can be made explicit by inverting 
Eqs.~(\ref{eq:Killing-energy})--(\ref{eq:Killing-ang-mom}), 
i.e., writing $dx^{b}/d \tau= g^{ba}U_{a}$ with the inverse metric $g^{ab}$ in the sector $(t,\newphi)$; then,
\begin{align}
   & \frac{dt}{d\tau}
   =
     - g^{tt} e + g^{t\newphi} \ell 
    = 
    [ f(r)]^{-1} \, \qty(e-\frac{J\ell}{2r^2})
    \; ,
    \label{eq:dtdtau1}
    \\
 &   \frac{d\newphi}{d\tau}
    =
 - g^{\newphi t} e + g^{\newphi \newphi} \ell
    =
    [ f(r)]^{-1} \, \qty(\frac{J e}{2r^2} -\frac{M\ell}{r^2}+\frac{\ell}{l^2})
    \; ,
\label{eq:dphidtau1}
\end{align}
In Eqs.~(\ref{eq:dtdtau1})--(\ref{eq:dphidtau1}), 
the metric tensor elements of Eq.~(\ref{eq:metric-BTZ_exp}) are used to write the components: 
$g^{tt}=g_{\newphi \newphi}/D= -1/f(r)$, 
$g^{\newphi \newphi}=g_{tt}/D =1/r^2 - J^2/[ 4r^4 f(r) ]$, 
and $g^{t \newphi}=- g_{t \newphi}/D= -J/[ 2r^2 f(r)]$, 
with the determinant $D  = g_{tt} g_{\newphi\newphi} - g_{t\newphi}^2 = - r^2 f(r) $ 
of the induced metric $g_{ab}$.

Equations~(\ref{eq:dtdtau1})--(\ref{eq:dphidtau1})
can also be derived and interpreted most easily in the local corotating frame
of Eqs.~(\ref{eq:metric-BTZ})--(\ref{eq:local_ang-velocity}), with angular velocity $\varpi$, where they have diagonal forms. Specifically,
with the formal local-frame transformation and reparametrization
\begin{equation}
\begin{aligned}
d\check{\newphi} &\equiv d\newphi - \varpi(r)\,dt
\\
\check{e}(r) &= e - \varpi(r)\ell
\end{aligned}
\label{eq:local-corotating}
\; .
\end{equation}
It should be noted that $\check{e}(r) $ is energy-like but actually a function of $r$, as it
does not have an associated Killing vector.
However, it plays the role of a local corotating energy associated with $\varpi$; and, in the near-horizon limit, it becomes the well-defined 
energy (per unit mass) with respect to the frame rotating with the black hole,
\begin{equation}
\begin{aligned}
\tilde{e} = \lim_{r \rightarrow r_{+}} \check{e} =
e -\Omega_{H} \ell
\label{eq:corotating-energy}
\end{aligned}
\; 
\end{equation}
which is the conserved time component of the momentum 
\begin{align}  
    &
    \tilde{e} = - \boldsymbol{\xi}_{(\tilde{t})} \cdot \boldsymbol{U}
= -U_{\tilde{t}} > 0
\; ,
      \label{eq:Killing-energy-tilde} 
\end{align}
associated with the Killing vector
$\boldsymbol{\xi}_{(\tilde{t})} $ of Eq.~(\ref{eq:Killing-vectors_corot-time})
via the black-hole corotating coordinates~(\ref{eq:rotating-coords_separation_1}).

Then, 
Eqs.~(\ref{eq:dtdtau1})--(\ref{eq:dphidtau1}) can be conveniently rewritten as
\begin{align}
& \frac{dt}{d\tau}=  [ f(r)]^{-1} \, \check{e}(r)
   \label{eq:dtdtau2}
\\
& 
 \frac{d\check{\newphi}}{d\tau}
    =
 \frac{1}{r^2}
 \ell
    \label{eq:dphidtau2}
 \; .
\end{align}
Thus,
$dt/d\tau$ is proportional to the local corotating energy $\check{e}(r)$; and $d\newphi/d\tau$ in the rotating frame, 
shifted by $\varpi$, in the form 
$ d\check{\newphi}/{d\tau} =  d{\newphi}/{d\tau}- \varpi  d t/{d\tau}$,
 is proportional to $\ell$.
(The coefficients are the inverse metric elements with respect to $t$ and $\check{\newphi}$.
See Appendix~\ref{app:BTZ-geometry-additional} for additional details.)
These equations are particularly useful for an efficient near-horizon framework, as shown in the next section.

Finally, a complete determination of geodesics requires $dr/d\tau$ in addition to
Eqs.~(\ref{eq:dtdtau1})--(\ref{eq:dphidtau1}); from the velocity normalization~(\ref{eq:velocity-normalization})
and Eq.~(\ref{eq:metric-BTZ_exp}), straightforward algebra gives
\begin{equation}
\qty(\frac{dr}{d\tau})^2  - \left( e - \frac{J\ell}{ 2 r^2} \right)^2 + f(r) \left( \frac{\ell^2}{r^2} + 1 \right) = 0
\; ,
\end{equation}
whence
\begin{equation}
\frac{dr}{d\tau}
=
\pm
    \qty[e^2+\frac{M\ell^2}{r^2}-\frac{e\ell J}{r^2}-\frac{\ell^2}{l^2}-f(r)]^{1/2}
    \; ,
\label{eq:drdtau1}
\end{equation}
where the signs $\pm$ correspond to outgoing and incoming trajectories respectively.
This radial geodesic equation can be rewritten in a more compact form in the local corotating frame,
 where it reads
\begin{equation}
\frac{dr}{d\tau}
=
\pm
    \qty[\check{e}^2
   - f(r) \left( \frac{\ell^2}{r^2} + 1 \right)
    ]^{1/2}
    \; ,
\label{eq:drdtau2}
\end{equation}
In conclusion, the geodesic equations, in their explicit forms of 
Eqs.~(\ref{eq:dtdtau1}), (\ref{eq:dphidtau1}), and (\ref{eq:drdtau1}), 
with compact corotating expressions given by
Eqs.~(\ref{eq:dtdtau2}), (\ref{eq:dphidtau2}), and (\ref{eq:drdtau2}), 
yield answers to 
all relevant questions about free fall motion. We will examine their near-horizon expressions and applications 
to HBAR in Secs.~\ref{sec:CQM_KGeq-geodesics} and \ref{sec:HBAR-thermodynamics}.

 An important application of the geodesic equations involves understanding the unusual properties of motions 
 in the ergosphere. One simple test of rotational behavior is provided by zero-angular-momentum ($\ell=0$) 
 particles or observers (ZAMOs)~\cite{frolov}.
This is best described from the viewpoint of a distant observer with the Killing time $t$, 
in the form of an angular velocity 
 $\Omega = \left.  \left( d\newphi/dt  \right) \right|_{\ell =0} $; 
 then, from either Eq.~(\ref{eq:Killing-ang-mom}) or
 Eqs.~(\ref{eq:dtdtau1}) and (\ref{eq:dphidtau1}), 
\begin{equation}
\Omega = 
 \left.  \frac{d\newphi}{dt}  \right|_{\ell =0} 
 = \left. \frac{d\newphi/d\tau}{dt/d\tau}   \right|_{\ell =0}  =
- \frac{ g_{ t \newphi } }{ g_{\newphi\newphi} } = \varpi =
\frac{J}{2r^2}
\; .
\label{eq:Omega-ZAMOs}
\end{equation}
Therefore, ZAMOs have an angular velocity~(\ref{eq:Omega-ZAMOs}) that exactly tracks 
the frame-dragging angular velocity of Eqs.~(\ref{eq:local_ang-velocity}) and ~(\ref{eq:local_ang-velocity_geom});
then, these are also ``locally nonrotating observers'' that define a local frame~\cite{MTW-gravitation, frolov}
 where the metric is diagonal.
These results give a complete operational characterization of the frame dragging formally given from 
the metric of Eqs.~(\ref{eq:metric-BTZ})--(\ref{eq:local_ang-velocity}).
Moreover, from the metric condition $ds^2 \leq 0$, physical geodesics
with $r= \text{constant}$ satisfy the inequality $\Omega_{-} \leq \Omega \leq \Omega_{+}$,
with the bounds $ \Omega_{\pm}= \varpi \pm \sqrt{\varpi^2 - g_{tt}/g_{\newphi \newphi}} $ 
corresponding to null trajectories. 
Inside the ergosphere, $g_{tt} >0$ implies that both $\Omega_{\pm}>0$,
forcing all the physical motions to follow the direction of the black hole rotation.
This proves the nonexistence of static solutions in the ergosphere, with the ergosurface 
 $g_{tt}=0$ as boundary where $\Omega_{-}=0$.

\section{BTZ Spacetime: Field Theory}
\label{sec:BTZ-spacetime_FT}

For black hole thermodynamics and the physics of HBAR, the behavior of quantum fields in a given spacetime 
geometry is the framework needed for all the relevant physical predictions.
For the sake of computational simplicity, it is customary to formulate the HBAR problem in terms of a massive scalar field whose quantum excitations are scalar (spin-0) quanta. Because the setup is motivated by quantum optics, these excitations are sometimes referred to as scalar ``photons,'' by analogy with ordinary vector (spin-1) photons~\cite{Scully_2018_HBAR, Ben-Benjamin-etal_2019_Unruh-rev}. Strictly speaking, however, the field excitations considered here are massive scalar quanta. The mass term is retained in the complete field equation but becomes subleading in the near-horizon expansion and therefore does not enter the leading conformal quantum mechanics dynamics. This scalar-field description captures the essence of the relevant phenomena, as has been established for a variety of related physical problems involving acceleration and gravitational backgrounds~\cite{AR_Scully-2003, AR_Belyanin-2006, Crispino-et-al_2008_Unruh-effect}, and allows for an efficient use of quantum-optics techniques adapted to scalar fields.

\subsection{Field theory in BTZ geometry}
\label{sec:FT-BTZ-geom}

The action for a real scalar field $\Phi(x)$ coupled
to a generic background geometry is 
\begin{equation}
           S[\Phi] 
    = \int d^3x
    \, \mathcal{L}(\Phi)
    = -
\frac{1}{2}
\int
d^{3} x
\,
\sqrt{-g}
\,
\left[
g^{\mu \nu}
\,
\nabla_{\mu} \Phi
\, 
\nabla_{\nu} \Phi
+ 
\mu_{\Phi}^{2} \Phi^{2}
+  \xi R \Phi^{2}
\right]
\; ,
\label{eq:scalar_action}
\end{equation}
where 
$\mu_\Phi$ is the scalar-field mass (with inverse length dimensions), 
$R$ the Ricci scalar curvature,
and $\xi$ is the dimensionless curvature coupling parameter.
By applying the action principle $\delta S[\Phi]/\delta \Phi = 0$ at the classical level, the Euler-Lagrange equations 
for the action~(\ref{eq:scalar_action}) give the generic Klein-Gordon equation for the real scalar field,
\begin{equation}
\left[ \Box - \left( \mu_\Phi^{2} + \xi R
 \right)  \right]\Phi 
\equiv 
\frac{1}{ \sqrt{-g} }\partial_{\mu} \left(\sqrt{-g} \,g^{\mu \nu}\,\partial_{\nu} \Phi\right)- \tilde{\mu}^{2} \Phi= 0
\; ,
\label{eq:Klein_Gordon_basic}
\end{equation}
where $\tilde{\mu}^2=\mu_\Phi^2+\xi R$. 
For the BTZ geometry, the metric tensor $g_{\mu\nu}$ is given in Eq.~(\ref{eq:metric-BTZ_exp}) and
the Ricci scalar takes the value $R = 6\Lambda = -6/l^2$; correspondingly, the 
Klein-Gordon equation~(\ref{eq:Klein_Gordon_basic}) reads
\begin{equation}
     \qty[  \frac{1}{r}\partial_r     \bigl( r f \partial_r \bigr)   
   +\qty(  g^{tt} \partial^2_t  +2g^{t\newphi}   \partial_t\partial_\newphi + g^{\newphi\newphi}\partial^2_\newphi )
    -\tilde{\mu}^2]
   \Phi(t,r,\newphi)=0
   \; ,
   \label{eomrtl2}
\end{equation}
where the reduced d'Alembertian in the sector $(t,\phi)$ is
\begin{equation} 
   g^{tt} \partial^2_t  +2g^{t\newphi}   \partial_t\partial_\newphi + g^{\newphi\newphi}\partial^2_\newphi 
   =
   -\frac{1}{f} \partial^2_t  
   -\frac{1}{f} \frac{J}{r^2} \partial_t\partial_\newphi 
   +\left( \frac{1}{r^2} -  \frac{J}{4 f r^2} \right) \partial^2_\newphi 
   \; .
\label{eq:reduced_dAlembert}
\end{equation}
In terms of the coordinates $(t,r,\newphi)$,
separation of variables
is allowed by the $(t,\newphi)$ independence of the coefficients. 
This is enforced with field modes of the form
\begin{equation}
    \phi_{m \omega}(t,r,\newphi) = R_{m \omega}(r)e^{-i\omega t}e^{i m \newphi}
    \; ,
 \label{fmnw1}
\end{equation}
where $m$ is the azimuthal angular momentum quantum number, which 
is restricted to values $m \in \mathbb{Z}$ due to the $2\pi$-periodicity of the angle $\newphi$
(corresponding to the angular momentum eigenvalue of $-i\partial_\newphi$). 
Here, the building blocks $e^{-i\omega t}$ and $e^{i m \newphi}$ can 
be chosen for the separated solution because:
 (i) they are complete sets of
functions with respect to the coordinates $t$ and $\newphi$; 
(ii) they are determined by the metric invariance 
with respect to $t$ and $\newphi$ using the Killing vectors of 
Eq.~(\ref{eq:Killing-vectors_stationary}) via
\begin{equation}
 \boldsymbol{\xi}_{({t})}   \phi_{m \omega} 
= -i \omega  \phi_{m \omega} 
\quad  
\text{and} \quad
  \boldsymbol{\xi}_{(\newphi)} \phi_{m \omega}
= i m \phi_{m \omega} 
\; .
\end{equation}
Correspondingly, the separated radial part of the mode inherits the mode numbers 
${\boldsymbol{s}} = (m, \omega)$, and 
straightforward algebra reduces Eqs.~\eqref{eomrtl2}--\eqref{eq:reduced_dAlembert} to
\begin{equation}
\qty[\frac{1}{r} \frac{d}{dr}\qty(r f(r) \frac{d}{dr})
+ \frac{1}{f(r)} \bigl(\omega - m \varpi \bigr)^2 - \frac{m^2}{r^2} - \tilde{\mu}^2] R_{m\omega}(r) = 0
 \; ,
\label{eomrmw}
\end{equation}  
using the frame-dragging angular velocity $ \varpi (r) = {J}/{2r^2}$
of Eqs.~(\ref{eq:local_ang-velocity}) and ~(\ref{eq:local_ang-velocity_geom}).
The resulting Eq.~(\ref{eomrmw}) can be more directly derived and interpreted 
in terms of a formal coordinate transformation, with new coordinates $\phi \rightarrow \phi - \varpi t$ and
$\omega \rightarrow \omega - m \varpi$, via the equivalent local factorization
$ \displaystyle \phi_{m \omega}(t,r,\newphi) = R_{m \omega}(r)e^{-i(\omega- m\varpi) t}e^{i m (\newphi - \varpi t)}$. 
This is indeed the field-mode equation in the locally rotating frame; in this form, the near-horizon analysis of the next section becomes straightforward.

 \subsection{Field quantization in BTZ geometry}
\label{sec:FT-BTZ-geom_quantization}

 In what follows, we will consider a complete set of modes 
$\bigl\{  \phi_{\boldsymbol{s}} (t,\mathbf{r}), \phi^{*}_{\boldsymbol{s}} (t,\mathbf{r}) \bigr\}$
as classical solutions to Eqs.~(\ref{eq:Klein_Gordon_basic}) or (\ref{eomrtl2}). 
The quantization of the theory involves the mode expansion of 
 quantum field operator $\Phi$ in the form~\cite{birrell-davies}
\begin{equation}
    \Phi(t, \mathbf{r}) 
    = \sum_{\boldsymbol{s}} \left[ a_{\boldsymbol{s}} 
     \phi_{\boldsymbol{s}} (t,\mathbf{r})
     + \mathrm{H.c.} \right]
    \; .    \label{eq:field_expansion}
        \end{equation}
        In Eq.~(\ref{eq:field_expansion}), $\mathrm{H.c.}$ is the Hermitian conjugate,
$(t, \mathbf{r})= (t, r, \newphi)$ stand for the spacetime coordinates,
 and the field modes $\phi_{\boldsymbol{s}} $ are identified by  
 the mode frequency $\omega$ and quantum numbers
 (collectively labeled by the symbol ${\boldsymbol{s}}$). 
 Thus, explicitly, with ${\boldsymbol{s}}= (m, \omega)$, 
\begin{equation}
    \Phi (t,r,\newphi) = 
    \sum_{m,\omega}\left[{a}_{m\omega}\phi_{m\omega}(t,r,\newphi) 
    + {a}^\dagger_{m\omega}\phi^*_{m\omega}(t,r,\newphi)\right],
\end{equation}
where ${a}^\dagger_{m\omega}$ and ${a}_{m\omega}$ are creation and annihilation operators, respectively. 
In addition, the modes are assumed to satisfy the orthonormality conditions
 \begin{equation}
 (\phi_{\boldsymbol{s}},  \phi_{\boldsymbol{s'}} ) 
 = - (\phi^{*}_{\boldsymbol{s}},  \phi^{*}_{\boldsymbol{s'}} ) 
 =   \delta_{ {\boldsymbol{s}}, {\boldsymbol{s'}} }
   \; \; , \; \; \; 
    (\phi^{*}_{\boldsymbol{s}},  \phi_{\boldsymbol{s'}} ) 
 = (\phi_{\boldsymbol{s}},  \phi^{*}_{\boldsymbol{s'}} ) 
   = 0
\; , 
   \label{eq:KG-orthonormality}
   \end{equation}
where the standard inner product~\cite{Crispino-et-al_2008_Unruh-effect, Takagi:1986},
\begin{equation}
(\Phi_1,\Phi_2) 
= 
 i \int_\Sigma 
\left(  \Phi_1^{*} \partial_\mu \Phi_2  -  \Phi_1 \partial_\mu \Phi_2^{*}  \right)
 d\Sigma^\mu
 \; ,
 \label{eq:KG-inner-product}
\end{equation}
is consistent with the Klein-Gordon equation.
In Eq.~(\ref{eq:KG-inner-product}), the integral is performed on a two-dimensional 
spacelike hypersurface $\Sigma$,
 with ``volume'' element $d \Sigma^{\mu} = n^{\mu} \sqrt{\gamma}\,  d^{2} x$ 
 oriented with the normal, future-directed ``time direction'' $n^{\mu}$ 
 (and a corresponding induced metric $\gamma_{ij}$). 
  These results (adjusted to the appropriate dimensionality) are valid in any number $D$ 
  of spacetime dimensions: the product is independent of the hypersurface $\Sigma$
according to Gauss's divergence theorem.
For example, in the polar coordinates used in the BTZ metric of 
 Eq.~(\ref{eq:metric-BTZ_exp}), leading to the field 
 equations~(\ref{eomrtl2})--(\ref{eq:reduced_dAlembert}),
 with $0\leq\newphi<2\pi$ and $r_+<r<\infty$, 
 the inner product reads
 \begin{equation}
(\Phi_1,\Phi_2) 
= 
i\int_\Sigma 
    d\newphi dr\sqrt{-g} (-g^{t\mu})\qty[\Phi_1^*(x)\partial_\mu \Phi_2(x)
    -\partial_\mu \Phi_1^*(x) \Phi_2(x)]
    \; ,
    \label{kginpd}
\end{equation}
where $\sqrt{\gamma} = \sqrt{-g} \sqrt{-g^{tt}(r)}$, $n^\mu = -g^{t\mu}(r)/\sqrt{-g^{tt}(r)}$, and
$d^2 x = dr\,d\newphi$; it is straightforward to verify for BTZ spacetimes in these coordinates that this expression 
has all the desired properties of an inner product, provided that appropriate regularization at the coordinate-singular
points is used~\cite{kenmoku2008normal}.
 
For implementation of the canonical quantization of the theory, the conjugate field momentum 
 is defined by
\begin{align}
\Pi(t,r,\newphi) &
= \frac{\partial\mathscr{L}}{\partial (\partial_t\Phi)} = -\sqrt{-g}g^{t\mu}\partial_\mu\Phi = -\sqrt{-g}g^{tt}\left[\partial_t + \varpi\partial_\newphi\right]\Phi 
\nonumber \\
& = i\sqrt{-g}g^{tt}\sum_{m,\omega}\bigl(\omega - m \varpi \bigr)\left[a_{m\omega}\phi_{m\omega}(t,r,\newphi) - a^\dagger_{m\omega}\phi^*_{m\omega}(t,r,\newphi)\right].
\end{align}
Quantization is explicitly enforced by the equal-time commutation relations,
\begin{align}
&    \left[{\Phi}(t,r,\newphi), {\Pi}(t,r',\newphi')\right] = i
 \delta^{(2)}(\mathbf{r}-\mathbf{r}')
\; , \label{feqcr1} \\
  &
    \left[{\Phi}(t,r,\newphi), {\Phi}(t,r',\newphi')\right] = 0 \; \; , \; \; \; \; 
     \left[{\Pi}(t,r,\newphi), {\Pi}(t,r',\newphi')\right] = 0 
    \; ,
    \label{feqcr2}
\end{align}
with
$ \delta^{(2)}(\mathbf{r}-\mathbf{r}')
= \delta(r-r')\delta(\newphi-\newphi')/r$.
The commutation relations in Eqs.~\eqref{feqcr1} and \eqref{feqcr2}
are enforced by the associated creation and annihilation commutators
\begin{eqnarray}
\left[{a}_{m\omega}, {a}_{\omega^\prime m'}^\dagger\right] 
= \delta_{\omega \omega^\prime}\delta_{m m'}
 \; \; , \; \; \; \; 
\bigl[ {a}_{m\omega} , {a}_{\omega^\prime m'} \bigr] = 0
 \; \; , \; \; \; \; 
 \bigl[{a}^\dagger_{m\omega}, {a}_{\omega^\prime m'}^\dagger\bigr] = 0
\; .
\end{eqnarray}
The vacuum state is defined by the set of annihilation statements
\begin{equation}
    {a}_{m\omega}\ket{0} = 0
 \; \; , \; \; \; \; 
 \text{for all} 
  \;  
 {\boldsymbol{s}}= (m, \omega)
 \; ,
\end{equation}
from which all other states can be generated by repeated application of creation operators.

\section{Conformal Quantum Mechanics (CQM) in BTZ Geometry: 
Near-Horizon Field Equations and Geodesics}
\label{sec:CQM_KGeq-geodesics}

In this section, we examine the near-horizon behavior of the theory.
This includes two sectors, which we analyze in this order:
(i)
 the field mode equations, highlighting their leading conformal quantum mechanics (CQM) behavior;
 (ii) the geodesic equations.
The relationship of this governing behavior to other metrics and dimensionalities is discussed,
stressing the universality of these findings. 

\subsection{Near-horizon field modes in BTZ geometry: CQM behavior}
\label{sec:nh-approx}

The near-horizon framework involves an approximation 
near the outer horizon
 \[
{\mathcal H}\equiv {\mathcal H}^{+}
\; \; , \; \;   r \sim r_{+}
\; ,
\]
with $r=r_{+}$ being the largest root of the scale-factor equation $f(r)= g^{rr}(r)= 0$, 
as described in Eq.~(\ref{eq:BH_horizons}) and the paragraph therein.
This scale factor is parametrized by the black hole mass $M$ 
and angular momentum $J$, as shown in Eqs.~(\ref{eq:BH_horizons}) and (\ref{eq:M-J_from_r-pm}).

 A hierarchical near-horizon approach can be defined with the Taylor series for the scale factor $f(r)$.
 The notation $\stackrel{(\mathcal H)}{\sim}$ will be used to represent this hierarchical expansion.
 By definition, the zeroth order vanishes: $f(r_{+})=0$.
 Thus, with the 
\begin{equation}
\text{shifted variable:} \; \; \; x= r -r_{+}
\; 
\label{eq:shifted-radial-coord}
\end{equation}
(radial coordinate from the outer horizon $r_+$), 
  the Taylor series starts at first or higher orders. 
 Considering the physically relevant {\em nonextremal\/} metrics, which satisfy the condition 
  $f'_{+} \equiv f'(r_{+}) \neq 0$, Eq.~(\ref{eq:f-prime}),
  the function $f(r)$ and its derivatives to second order are  
\begin{equation}
f(r)  \stackrel{(\mathcal H)}{\sim}  f'_{+}  \, x \left[ 1 +
\mathcal{O}(x) \right]  \, , \; \; \; 
f'(r)  \stackrel{(\mathcal H)}{\sim}  f'_{+} \left[ 1 + \mathcal{O}(x) \right]   \, , \; \; \; 
f''(r)  \stackrel{(\mathcal H)}{\sim}  f''_{+} \left[ 1 + \mathcal{O}(x) \right] \, ,
\label{eq:nh-expansions}
\end{equation}
where $f'_{+}$ is given by Eq.~(\ref{eq:f-prime}) and
\begin{equation}
f''_{+} \equiv f''(r_{+}) 
= \frac{2}{l^2} 
\left[
1 + 3 \left( \frac{r_{-}}{r_{+}} \right)^2
\right]
=
\frac{2}{l^2} 
+\frac{3}{2} \frac{J^2}{r_+^4}
\label{eq:f-second-prime}
\; .
\end{equation}
This near-horizon approximation can be applied to the 
original Klein-Gordon equation~\eqref{eomrmw}.
This equation involves the differential operator
$\Delta_r$ as the radial part of the d'Alembertian,
\begin{equation}
    \Delta_r = \frac{1}{r} \frac{d}{dr} 
    \left( r f(r)\frac{d}{dr} \right)
    \label{eq:radial-operator}
    \; ,
\end{equation}
with leading behavior $\mathcal{O}(x^{-1})$ (as defined by scaling):
\begin{equation}
    \Delta_r  
    \stackrel{(\mathcal H)}{\sim}
    \frac{f'_{+}}{r_{+}} \frac{d}{d  x} \left( x  \frac{d}{d x} \right) 
    \label{eq:radial-operator_exp}
    \; .
\end{equation}
The other, non-differential, terms are $\mathcal{O}(1)$ and higher orders, except for the
leading divergence in the frequency term,
\begin{equation}
 \frac{1}{f(r)} \bigl(\omega - m \varpi \bigr)^2 
 \stackrel{(\mathcal H)}{\sim}
  \frac{1}{f'_{+} x} \bigl(\omega - m \Omega_{H})^2 
  \; ,
  \label{eq:frequency-term_exp}
 \end{equation}
 where $ \Omega_{H}$ is the black hole angular velocity defined in terms of its outer horizon,
 Eq.~(\ref{eq:BH_ang-velocity}). Here, both the scale factor $f(r)$ and frame-dragging velocity $\varpi (r)$ involve higher-order terms that would only contribute at  $\mathcal{O}(1)$ and higher orders of the overall expansion.
With Eqs.~(\ref{eq:radial-operator_exp}) and (\ref{eq:frequency-term_exp}), 
the leading-order of the field equation becomes
\begin{equation}
\left[\frac{1}{x} \frac{d}{d  x} \left( x  \frac{d}{d x} \right) 
+ \left( \frac{ {\tilde{\omega}}}{f'_{+}} \right)^{2}
\frac{1}{x^2} \right] R(x)
\stackrel{(\mathcal H)}{\sim} 0 \; ,
\label{eq:Kerr_Klein_Gordon_conformal-R}
\end{equation}
where
\begin{equation}
    \tilde{\omega}=\omega - m\Omega_{H}
    \label{crfw-noh1}
\end{equation}
is the corotating frequency associated with the horizon angular velocity.

Equation~\eqref{eq:Kerr_Klein_Gordon_conformal-R}
is manifestly scale invariant, as all of its terms have $\mathcal{O}(x^{-2})$
scaling. This is one of the standard forms of conformal quantum mechanics (CQM). Its simplest representation
can be obtained via reduction to its normal form (``Schr\"{o}dinger-like'') 
with the Liouville transformation $ R(x) \sim x^{-1/2} u(x) $:
\begin{equation}
    u''(x) + \frac{\lambda}{x^2} \left[1 + \mathcal{O}(x)\right] u(x) = 0,
\label{eq:Klein_Gordon_conformal}
\end{equation}
where the coupling constant $\lambda$ encodes the effective-potential spectral properties
\begin{equation}
    \lambda = \frac{1}{4} + \Theta^2
    \quad , \quad
     \Theta = \frac{\tilde{\omega}}{f'_{+}} 
     \equiv \frac{\tilde{\omega}}{2\kappa} 
          \label{eq:conformal_interaction}
    \; ;
\end{equation}
via the frequency-dependent ``conformal coupling'' parameter $\Theta$,
with the BTZ surface gravity $\kappa$ of Eq.~(\ref{eq:surface-gravity_BTZ}).

\subsection{CQM symmetry and field modes}

 Equation~\eqref{eq:Klein_Gordon_conformal} can be analyzed as an effective Hamiltonian problem that defines conformal quantum mechanics (CQM) as a $(0+1)$-dimensional field theory with conformal SO(2,1) symmetry. Specifically, 
the near-horizon regime emerges as an effective theory that captures the essence of black-hole thermodynamics 
and the HBAR physics of particles falling into a black hole, via
Eqs.~(\ref{eq:Kerr_Klein_Gordon_conformal-R})--(\ref{eq:conformal_interaction}).
The one-dimensional effective Hamiltonian $\mathscr{H}$ in
 Eq.~(\ref{eq:Klein_Gordon_conformal}), i.e.,
  \begin{equation}
  \mathscr{H} = {p}_{x}^{2}-  \frac{ \lambda }{ x^{2} }
  \; ,
  \label{eq:CQM-nh-Hamiltonian}
  \end{equation}
 governs the dominant near-horizon physics, and is a particular realization of
 the inverse square potential of conformal quantum mechanics~\cite{Qanomaly-molecular-to-BH}.

 The near-horizon physics exhibits an {\em asymptotic conformal symmetry\/}.
This is described by the CQM framework based on the inverse-square-potential Hamiltonian $\mathscr{H}$ of Eq.~(\ref{eq:CQM-nh-Hamiltonian}), which is manifestly scale invariant 
 as it is homogeneous of degree $-2$~\cite{Cam_DT1,Cam_DT2}. Moreover,
this operator is part of an SO(2,1) symmetry group
associated with a lower-dimensional conformal field theory~\cite{dAFF,jackiw1,jackiw2,jackiw3}.
The algebra of this SO(2,1) group consists of three operators:
$\mathscr{H} $, as the generator of time displacements (with respect to 
a specified time $t$
conjugate to
$\mathscr{H} $),
along with
the dilation operator
$
D= t H - 
\left( x p + p x \right)/4
 $
 (where $p$ is the conjugate momentum),
 performing scale transformations,
and the special conformal operator
$
K= t^{2} H -  
t \, (px + x p)/2
 + 
x^{2}/4
$
as the generator of inverse time displacements.
These operators generate the 
SO(2,1) {\em conformal algebra\/}
$
[D,H]
= - i \hbar H
  \;  ,
\; \;
[K,H]
= - 2 i \hbar D
\;  ,
\; \;
[D, K]
=  i \hbar K
$, which has been recently studied in its most general setting using path integral 
methods~\cite{Cam_CQM-PI}, extending earlier results on path integrals 
of the inverse square potential~\cite{Cam_CQM-GF-ISP, Cam_SQM-PI}.

The solutions to Eq.~\eqref{eq:Klein_Gordon_conformal} take the form
\begin{equation}  
    u_{\pm}(x) = x^{1/2 \pm i\Theta} = \sqrt{x} e^{\pm i\Theta \ln x}
    \; ,
    \label{ctheta12}
\end{equation}  
where the $(+)$ and $(-)$ signs denote outgoing and ingoing waves, respectively,
normalized as asymptotically exact WKB local waves~\cite{nhcamblong-sc}.
The characteristic logarithmic phase dependence of these solutions reveals emergent scale invariance, a definitive signature of underlying conformal symmetry near the horizon.
Furthermore, 
in the standard treatment of CQM in the strong-coupling sector, the conformal parameter is positive: 
$\Theta=
\tilde{\omega}/{2\kappa} 
>0$. Here,
this is due to the condition
\begin{equation}
 \tilde{\omega} > 0
 \; 
 \end{equation}
  on the corotating frequency given in Eq. \eqref{crfw-noh1}, which is
 required for the canonical quantization with positive frequency modes defined with respect 
 to the Killing vector $ \boldsymbol{\xi}_{(\tilde{t})} \equiv \partial_{\tilde{t}}$ of
 Eq.~(\ref{eq:Killing-vectors_corot-time}):
 \begin{equation}  
 \boldsymbol{\xi}_{(\tilde{t})}
  \phi_{m\omega} = -i\tilde{\omega} \phi_{m\omega}
    \; ,  
\end{equation} 
 see, e.g., \cite{kuwata2008eigenvalue}. 
 In fact, this is identical to the treatment of all problems with an angular velocity, 
 as familiar from the Kerr spacetime quantization~\cite{acceler-rad-Kerr}.
 In this setting, associated with corotating coordinates, the corotating azimuthal angle
 $\tilde{\newphi}$ defines an angular velocity
$\displaystyle
    \tilde{\Omega}(r) = d\tilde{\newphi}/{dt} =d(\newphi-\Omega_{H} t)/{dt}= \Omega(r) - \Omega_{H} $
    that vanishes at the outer horizon---the near-horizon physics is effectively static in the corotating frame.
 In essence, this argument shows that,
   \begin{quotation}
   \noindent
   with the squared-lapsed factor $f(r)$ of Eq.~(\ref{eq:f-factor}),
  {\it the near-horizon BTZ metric is a generalized Schwarzschild metric
   in the corotating frame.\/}
  \end{quotation}

 Finally, the complete expression of the CQM modes is obtained with the proportionality
 $R(x) \sim x^{-1/2} u(x)$ for the radial function, along with the temporal-angular factor rewritten 
 with the factorization 
  $e^{-i\omega t} e^{i m \newphi} = e^{-i\tilde{\omega} t} e^{i m \tilde{\newphi}}$, where $\tilde{\omega}$ is the corotating frequency given in Eq. \eqref{crfw-noh1}, and $\tilde{\newphi} = \newphi - \Omega_{H} t$ is the corotating azimuthal angle. As a result,
the outgoing and ingoing near-horizon CQM field modes are
\begin{equation}
    \phi_{m\omega}(t,r,\newphi)
    \stackrel{(\mathcal H)}{\sim} 
     \Phi^{\pm {\rm \scriptscriptstyle (CQM)}}_{m\tilde{\omega}}(t,x,\tilde{\newphi})
         \stackrel{(\mathcal H)}{\propto} 
          e^{\pm i\Theta \ln(x)} e^{-i\tilde{\omega} t} e^{i m \tilde{\newphi}}
    \; .
    \label{eq:CQM_modes}
\end{equation}

\subsection{Near-horizon limit from exact field mode solutions}
\label{sec:nh-limit_from_exact-solution}

Remarkably, the scalar field mode equations admit an exact solution for the BTZ geometry.
This allows for an alternative derivation of the CQM field modes~(\ref{eq:CQM_modes}) within a global
perspective.

The exact solution can be obtained from 
the radial differential equation~\eqref{eomrmw}, which is written
 for the radial function $R_{m\omega}(r)$.
As discussed in Appendix~\ref{app:exact-solution_scalar-BTZ},
this equation has three regular singular points:
(i) the inner horizon $r=r_{-}$;
(ii) the outer horizon $r=r_{+}$;
and
(iii) spatial infinity $r = \infty$.
As a result,
it can be recast in the form of
a hypergeometric differential equation~(\ref{eq:hypergeometric-DE})
  for a function $f_{m\omega}(u)$,
via an appropriate transformation of variables, such that 
the singular points are mapped to
 $u = 0$, $u = 1$, and $u = \infty$.
 It should noted that we are using a variable $u$ that is unrelated to the use of $u(r)$ as reduced radial
 mode in the previous section---this should not pose any ambiguity for the remainder of the paper.

 The general solution to Eq.~(\ref{eq:hypergeometric-DE})
 is built from two linearly independent functions 
proportional to a Gaussian hypergeometric function (series) 
${}_{2}F_{1}(a,b;c;z)  \equiv F(a,b;c;z) $,
with parameters $(a,b,c)$,
including the whole family of such functions within the set of 
Kummer's 24 solutions~\cite{Abramowitz-Stegun-1972}.
  In this article, we use the transformation~\cite{BTZ-Tmodes-Eucli_Ichinose-1995, ortiz2012no} 
   \begin{align}
   & u=  \frac{r^2-r^2_{-}}{r^2_{+} - r^2_{-}}
    \; ,
    \label{uveql1}
    \\
       &  R_{m\omega}(u)= 
        (u-1)^{\alpha_{+}} u^{\alpha_{-}} 
        f_{m\omega}(u)
          \label{eq:radial-DE-function-transf-u}
        \; .
\end{align}
In Eq.~(\ref{eq:radial-DE-function-transf-u}),
the constants $\alpha_{\pm}$
are purely imaginary,
    \begin{equation}
    \alpha_{\pm} =  i \Theta_{\pm}
    \; , \; \; 
    \text{where}
    \; \; \; \;
    \Theta_{\pm} =
\frac{\omega - m \Omega_{\pm}
}{2 |\kappa_{\pm}| }
    \; ,
\label{eq:alpha_pm}
    \end{equation}
    where the parameters $\Omega_{\pm}$ and $\kappa_{\pm}$ are the angular velocity and surface gravity
    at the outer and inner horizons, respectively,
       \begin{equation}
\Omega_{\pm} =
\lim_{r \rightarrow r_{\pm}} \varpi
 =
 \frac{r_{\mp}}{r_{\pm} l}
 \; ,
 \;  \;  \;  \;
  \kappa_{\pm}= \frac{f'_{\pm}}{2} =
   \frac{r_{\pm}^2 - r_{\mp}^2}{ l^2 r_{\pm}}
\label{eq:ang-velocity_surface-gravity_both-H}
    \; ,
    \end{equation}
    generalizing Eqs.~(\ref{eq:BH_ang-velocity}) and (\ref{eq:surface-gravity_BTZ}) to 
    the known expressions for both horizons. For the outer horizon,
    $\Theta_{+} = \Theta$ is the conformal parameter~(\ref{eq:conformal_interaction})
     that governs the CQM physics of the near-horizon physics of field modes. 
     This shows that similar expressions relate to the inner horizon, and the corresponding physics is supported 
     by the hypergeometric differential equation~(\ref{eq:hypergeometric-DE})
     via its parameters 
     \begin{equation}
a=i \left( \Theta_{+} + \Theta_{-} \right) +\frac{1}{2} \Delta_{+}
\; , \; \; \; \;
b=i \left( \Theta_{+} + \Theta_{-} \right) + \frac{1}{2} \Delta_{-}
\; , \; \; \; \;
c=2 i \Theta_{-}+1
\; ,
    \label{abcdef}
\end{equation}  
where the constants      
   \begin{equation}
   \Delta_{\pm} = 1 \pm \sqrt{1+\tilde{\mu}^2l^2}      
  \label{eq:Delta_pm} 
  \end{equation} 
    are the scaling dimensions in the dual CFT$_{2}$ that couples to the 
    bulk scalar field~\cite{Klebanov_TASI_AdS/CFT};
      $\Delta \equiv \Delta_{+} $ is the standard conformal weight 
       in AdS$_{3}$/CFT$_{2}$ and  $\Delta_{-} \equiv \tilde{\Delta} = 2- \Delta $ is
       the complementary scaling dimension of the shadow 
       operator~\cite{Ferrara-etal_shadow-1972, simmons-Duffin_shadow-2014}.

Enforcing a boundary condition of the Dirichlet type
 at asymptotic spatial infinity selects one of the two orthogonal radial solutions 
 $R^{(1,2)}$
 with expansions around $u=\infty$;
  these are selected from Kummer's 24 solutions,
  with $z=1/u$:
\begin{equation}
    f_{m\omega}(u)
   = \bigl\{ 
    u^{-a}F(a,a-c+1;a-b+1;u^{-1})\;,\; u^{-b}F(b,b-c+1;b-a+1;u^{-1})
    \bigr\}
    \; 
    \label{eq:hypergeometric-solution_infty}
    \end{equation}
     (Eqs.~15.5.7 and 15.5.8
      in Ref.~\cite{Abramowitz-Stegun-1972}, with $z=u$).
    As shown in 
 Appendix~\ref{app:exact-solution_scalar-BTZ} or by knowledge of AdS properties,
 these solutions scale as power laws $R^{(1,2)}     \stackrel{(r \rightarrow \infty)}{\propto} 
  r^{-\Delta_{\pm}}$ with the conformal weights
$ {\Delta_{\pm}}$, which imply that only $R^{(1)}$ has the correct asymptotic behavior
due to ${\rm sgn}(\Delta_{\pm}) = \pm 1$.
Then, the physical field modes, including all the coordinates with Eq.~(\ref{fmnw1}), are
\begin{equation}
    \phi_{m\omega}(t,u,\newphi)
   =
    e^{-i\omega t}
    e^{im \newphi}
    (u-1)^{\alpha_{+}} u^{\alpha_{-} -a}
    F(a,a-c+1;a-b+1;u^{-1})
    \; .
    \label{nheq}
\end{equation}

 The all-important near-horizon expansions can be obtained with the solutions around $u=1$, which are 
chosen as the pair of Kummer solutions
    \begin{equation}
         \begin{aligned}
    R_{m\omega}(u)
    = &
    \bigl\{
   (u-1)^{\alpha_{+}} u^{\alpha_{-}} 
   F(a,b;a+b-c+1;1-u)
   \\
   &
      , 
     (u-1)^{-\alpha_{+}} u^{-\alpha_{-}} 
        F(1-b,1-a;c-a-b+1;1-u)
        \big\} 
         \stackrel{(\mathcal H)}{\propto} 
           \bigl\{
           x^{i\Theta},   x^{-i\Theta}
         \big\} 
         \; ,
\end{aligned}
  \label{eq:hypergeometric-solution_nh}
\end{equation}
where $ \alpha_{+} = i \Theta$ gives the correct exponents for the leading terms, as
$u-1   \stackrel{(\mathcal H)}{\propto}  2x/\kappa l^2$.
Thus, 
\begin{equation}
    \phi_{m\omega}(t,r,\newphi)
         \stackrel{(\mathcal H)}{\propto} 
          x^{\pm i\Theta} e^{-i\tilde{\omega} t} e^{i m \tilde{\newphi}}
    \; .
    \label{eq:CQM_modes_2}
\end{equation}
which gives exactly the CQM modes 
$  \Phi^{\pm {\rm \scriptscriptstyle (CQM)}}_{m\tilde{\omega}}(t,x,\tilde{\newphi})$
 of Eq.~(\ref{eq:CQM_modes}).

\subsection{Near-horizon BTZ geodesics}

We now turn to a discussion of relevant results for the near-horizon geodesics in the BTZ geometry. 
The exact BTZ geodesic equations, in their forms of 
Eqs.~(\ref{eq:dtdtau1}), (\ref{eq:dphidtau1}), and (\ref{eq:drdtau1}), yield answers to 
all questions about free fall motion. 
In taking the near-horizon limit, the more physically transparent and compact expressions of 
Eqs.~(\ref{eq:dtdtau2}), (\ref{eq:dphidtau2}), and (\ref{eq:drdtau2}) are used below.
Additional technical details are described in Appendix~\ref{app:BTZ-geodesics}. 

For the purpose of evaluating the probability amplitudes for atomic transitions of freely falling atoms, 
 the functional dependences $x^{\mu}(\tau)$ are needed. This corresponds to the 
 physics depicted in Fig.~\ref{fig:HBAR-setup}.
 The dynamics of the atoms is governed by their proper times $\tau$, and
 the goal is to determine the relationship between $r$ and $\tau$ in the near-horizon limit, leading to explicit expressions of all the coordinates in terms of $x=r-r_{+}$.
 Moreover, the relevant geodesics are ingoing, as atoms are freely falling into the black hole;
 thus, with the negative sign, 
  as proper time $\tau$ increases, the particle's radial coordinate $x$ decreases.
   Therefore, the first step is to determine the near-horizon limit of the
 ingoing radial geodesic~(\ref{eq:drdtau2}), which takes the form
 \begin{equation}
    \frac{d\tau}{dx}
    \stackrel{(\mathcal H)}{\sim}
   =
  - \left[\check{e}^2+ \mathcal{O}(x) \right]^{-1/2}
   \stackrel{(\mathcal H)}{\sim} \tilde{e}^{-1}
 + \mathcal{O}(x)
  \; ,
\label{dtdxb0b1}
\end{equation}
because, in the near-horizon limit, $\check{e}  \stackrel{(\mathcal H)}{\sim} \tilde{e}=e - \Omega_{H} \ell$
is the corotating energy defined in Eqs.~(\ref{eq:corotating-energy})--(\ref{eq:Killing-energy-tilde}).
Thus, 
this geodesic equation is integrated to 
  \begin{equation}
\tau \stackrel{(\mathcal H)}{\sim}
 -k  x+ \mo(x^2) + \text{const} 
\; ,
\label{eq:tau-vs-x}
\end{equation}
where $k=\tilde{e}^{-1}$. 
Of course, the integration constant can be removed by an arbitrary initial-time choice, and the higher orders
become negligible as the near-horizon region is approached.

As a next step,
the original radial geodesic~(\ref{eq:drdtau2}) is used in 
 Eqs.~(\ref{eq:dtdtau2}) and (\ref{eq:dphidtau2}), in conjunction with further use
 of the expansions~(\ref{eq:shifted-radial-coord})--(\ref{eq:f-second-prime})
  to give the integrated near-horizon forms of the time and azimuthal coordinates,
\begin{align}
    t  &\stackrel{(\mathcal H)}{\sim} - \frac{1}{2 \kappa} \ln x - C x +
    \mo(x^2) 
    \; , \label{eq:t-vs-x} \\
    \tilde{\phi} &\stackrel{(\mathcal H)}{\sim} \alpha x +\mo(x^2)\; . 
     \label{eq:phi-vs-x}
\end{align}
where the values of the constants $C$ and $\alpha$ are given in Appendix~\ref{app:BTZ-geodesics}. 
In the derivation of the emission and absorption probabilities (next section), 
 all the coordinates are kept through order $\mo(x) $, as this is the leading order
 for the proper time $\tau$. The linear terms in $x$ are
governed by the three parameters $k=\tilde{e}^{-1}$, $C$, and $\alpha$, which encode the initial conditions 
of the atoms via their specific corotating energy $\tilde{e}$ and angular momentum $\ell$.
Ultimately, these constants do not appear in 
dominant part of the HBAR acceleration transition probabilities, which 
are determined by the governing scale symmetry of CQM.
Instead, the metric spacetime symmetries only act as tools that simplify the geodesic initial value problem; with this
caveat, the results of Eqs.~(\ref{eq:t-vs-x}) and (\ref{eq:phi-vs-x}) 
are the geodesic equivalent of the hierarchical near-horizon expansion of the metric 
in Eq.~(\ref{eq:nh-expansions}).
In particular, the logarithmic term dominates when $x \rightarrow 0$, with its
characteristic scale invariance associated with 
the familiar gravitational frequency shift~\cite{parkerprl}. This is the same type of conformal invariance that yields the logarithmic phase of the field modes~(\ref{ctheta12}).

In conclusion, the basic near-horizon analysis shows that both the field modes and the geodesics are controlled
by the scale-invariant features of CQM.

\section{HBAR Thermodynamics in BTZ Geometry}
\label{sec:HBAR-thermodynamics}

\subsection{HBAR physics setup of black hole geometry and falling atoms}

In this article, we adopt the HBAR model of acceleration radiation in black holes of
Refs.~\cite{Scully_2018_HBAR, acceler-rad-Schwarzschild, acceler-rad-Kerr, acceler-rad-Qopt-1, acceler-rad-Qopt-2}. 
The HBAR model can be characterized via a thought experiment
 consisting of atoms that are randomly injected, following free-fall
 paths in the given gravitational background around a black hole.
This may be regarded as an analog of an optical cavity in a lab, 
simulating a system with boundary mirrors.
Specifically,
a cloud of two-level atoms is injected at random times, simulating
the analog of a cavity around the black hole, where one end of the cavity is on the horizon and the other end is far away from the injection point. The atoms then fall freely towards the black hole,
as in Fig.~\ref{fig:HBAR-setup}, while interacting with the background field via a dipole-type interaction, Eq.~(\ref{eq:QO_interaction_potential}).
Here, this ``quantum-optics setup'' yields a probe of the near-horizon black hole behavior, in a manner 
that requires appropriate boundary conditions for the quantum fields.

The quantum-optics setup is achieved with an initial configuration of the quantum field prepared 
in a Boulware vacuum around the black hole. 
This choice of a vacuum is motivated by the quantum-optics analog, 
and selects an initial state with Hawking radiation suppressed.
The boundary condition amounts to the presence of an ideal mirror held at the outer
horizon, shielding any radiation coming out of the black hole.
This gedanken experiment is explained in greater detail in the previous 
Refs.~\cite{Scully_2018_HBAR, acceler-rad-Schwarzschild, acceler-rad-Kerr}. 
The technicalities of the Boulware vacuum in the case of BTZ black hole are also discussed in Appendix~\ref{app:exact-solution_scalar-BTZ}. 
For the purpose of deriving HBAR properties, any Boulware-like state suffices, 
and we use this construction in this section.
(All Boulware-like states yield the same result for the relevant transition probabilities; 
see discussion in Ref.~\cite{acceler-rad-Kerr}.)

In short,
the imaginary cavity associated with the black hole should be viewed as an auxiliary construction within a thought experiment that simulates a cavityless black hole with a Boulware state---an actual mirror would probably be unstable under reasonable assumptions of causal sound wave propagation.
Moreover,
a Boulware vacuum would not be the physically reasonable assumption if one considered an ordinary black hole formed from the usual processes of gravitational collapse, as this system would normally end up with an Unruh vacuum instead~\cite{QBH-primer_Brout-etal_1995, BH-evap_Fabbri-NavarroSalas_2005}.
However, if caution is exercised in its interpretation, the proposed HBAR thought experiment remains a powerful 
device to probe the interplay between fundamental quantum effects and strong gravitational fields, and provides an alternative probe of black hole thermodynamics.

\subsection{Atom-field-gravity interactions}

The basic setup of the HBAR model treats the atom as a two-level system, in a form that captures the essence of the relevant physics.
This is a generalization of the quantum Rabi model of widespread use in quantum optics and quantum 
information~\cite{scullybook,meystrebook, QRM-rev_Braak-et-al-2016}.
In its simplest form, the atom's Hamiltonian (in units with $\hbar=1$) is given by 
$\mathscr{H}_{\rm at} = \left(\ket{a}\bra{a} - \ket{b}\bra{b}\right) \nu/2$,
 where $\ket{a}$ and $\ket{b}$ are its excited and ground states respectively, and $\nu$ is the atomic transition frequency. 
The total Hamiltonian of the field-atom system is
$H = H_{\rm at} + H_{\rm field} + H_{\rm int}$,
 where the field operator is expanded with a complete set of orthonormal modes 
  $ \phi_{\boldsymbol{s}} (\boldsymbol{r}, t) $ according to Eq.~(\ref{eq:field_expansion}),
  which implies that the
 quantum field Hamiltonian takes the form of a sum of harmonic-oscillator modes.
The interaction Hamiltonian ${H}_{\rm int} \equiv V_{I} $ is
given by
\begin{equation}
{H}_{\rm int} 
=
g \, \Phi ( {\boldsymbol{r }} (\tau), t(\tau) ) \, \sigma (\tau)
 \; ,
\label{eq:QO_interaction_potential}
\end{equation}
where $\sigma$ is the atomic-state transition operator defined below,
and both the field operator $\Phi$ and $\sigma$
are evaluated at the proper time $\tau$ of the atom. 
This is the monopole analog of a dipole coupling for a spin-one photon field;
the scale can be adjusted by considering
a coupling $g =\mu E/\hbar$, where $\mu$ is the atomic dipole moment and E is the electric field.
The atomic state transitions are described by the evolving linear-combination operators
\begin{equation}
\sigma (\tau) = \sigma_- e^{-i\nu \tau} + \sigma_+ e^{i\nu \tau}
=
\sigma_- e^{-i\nu \tau} + \mathrm{H.c.} 
\; , 
\label{eq:QO_atomic-transitions}
\end{equation}
where $\mathrm{H.c.}$ is the Hermitian conjugate and
$\sigma_{\mp}$ are the atomic lowering and raising operators
	${\sigma}_{-} \equiv {\sigma}_{ba} = \ket{b}\bra{a} $ and
	${\sigma}_{+} \equiv {\sigma}_{ab} = \ket{a}\bra{b} $
(lowering/raising Pauli matrices).

The interaction Hamiltonian~(\ref{eq:QO_interaction_potential}) 
in the interaction picture generally includes four terms, 
as follows from the expansions~(\ref{eq:field_expansion}) and (\ref{eq:QO_atomic-transitions})
 of the field and atomic operators.
The experiment is set up with an initial condition that the field is in a Boulware vacuum  $\ket{0}$
and the atoms are in their ground state $\ket{b}$. 
Then, the two relevant terms for emission and absorption of field modes involve a transition 
of the atom to its excited state $\ket{a}$ with the simultaneous field transition from $\ket{0}$ to the 
state $\ket{1_{\boldsymbol{s}}}$ with one photon or vice versa.
As a result, up to first order in perturbation theory, for the field mode ${\boldsymbol{s}}$,
the emission probability $P_{{\mathrm e}, {\boldsymbol{s}} }$ 
 is
\begin{equation}
P_{{\mathrm e}, {\boldsymbol{s}} } 
= \left| \int d\tau \;\bra{1_{\boldsymbol{s}},a}V_I(\tau)\ket{0,b}\right|^2 
  = g^2 \left|\int\; d\tau\; \phi_{\boldsymbol{s}}^*(\mathbf{r} (\tau),t(\tau)) \, e^{i\nu\tau}\right|^2
        \; ;
        \label{eq:P_ex_explicit}
\end{equation}
and the absorption probability is 
\begin{equation}
    P_{{\mathrm a}, {\boldsymbol{s}} }
     = \left|\int d\tau \;\bra{0,a}V_I(\tau)\ket{1_{\boldsymbol{s}},b}\right|^2
     = g^2 \left|\int\; d\tau\; \phi_{\boldsymbol{s}}(\mathbf{r} (\tau),t(\tau)) \, e^{i\nu\tau}\right|^2
        \; .
         \label{eq:P_ab_explicit} 
\end{equation}
     The probabilities in Eqs.~(\ref{eq:P_ex_explicit}) and (\ref{eq:P_ab_explicit}) 
     are the key expressions that lead to the detailed balance that characterizes the thermal properties
     of the radiation field in the HBAR thought experiment. For this calculation,
 the field modes $\phi_{\boldsymbol{s}}(\mathbf{r} (\tau),t(\tau))$ depend on the black hole geometry and the 
falling-atom trajectories given by the geodesic equations, as studied in the previous 
Secs.~\ref{sec:BTZ-spacetime} and \ref{sec:CQM_KGeq-geodesics}.

\subsection{BTZ probabilities: Conformal nature}
\label{sec:BTZ_conformal}

The emission and absorption probabilities of scalar photons by freely falling atoms in the BTZ geometry
follow from Eqs.~(\ref{eq:P_ex_explicit}) and (\ref{eq:P_ab_explicit})
with the 
 outgoing CQM modes~(\ref{eq:CQM_modes}), 
 via the assignments
   $\phi_{(+);{\boldsymbol{s}}} \sim
\left[ \Phi^{+ {\rm \scriptscriptstyle (CQM)}} \right]^{*}$
 and
    $\phi_{(-);{\boldsymbol{s}}} \sim
 \Phi^{+ {\rm \scriptscriptstyle (CQM)}}$; 
 then,
   \begin{equation}
   \left\{ \begin{array}{l}
  P_{{\mathrm e}, {\boldsymbol{s}} } \\
  P_{{\mathrm a}, {\boldsymbol{s}} }
  \end{array}\right\}
\stackrel{(\mathcal H)}{\sim}\; 
{g^2} k^2
\left|
\int_0^{x_f} \; dx\; 
x^ {\mp i\Theta} e^{ \mp i m \tilde{\phi}(x)} 
e^{ \pm i\tilde{\omega}t(x)} e^{i\nu\tau(x)} 
\right|^2
    \label{eq:P_CQM}
    \; ,
     \end{equation}
 where the integrals can be reformulated to leading order in terms of the near-horizon variable $x$ via 
 Eq.~(\ref{eq:tau-vs-x}), with 
 $\displaystyle   k=\frac{1}{\tilde{e}}$.
 
 In the actual thought experiment, 
 there is a beam of atoms falling into the black hole,
 prepared with an injection rate $\mathfrak{r}$; thus, 
the emission and absorption rate coefficients are
$R_{{\mathrm e}, {\boldsymbol{s}} }  =  \mathfrak{r} P_{{\mathrm e}, {\boldsymbol{s}} } $
and
$  R_{{\mathrm a}, {\boldsymbol{s}} } =   \mathfrak{r}  P_{{\mathrm a}, {\boldsymbol{s}} }$.
   In particular, when $\tau$, $t$, and $\tilde{\phi}$ are replaced in the integrand of Eq.~(\ref{eq:P_CQM}) with
  Eqs.~(\ref{eq:tau-vs-x}), (\ref{eq:t-vs-x}), and (\ref{eq:phi-vs-x}), the emission rate
  takes the form
\begin{equation}
        R_{e, {\boldsymbol{s}} }   =  \mathfrak{r} \, g^2 k^2 \left|\int_0^{x_f} 
        dx \;x^{-i\tilde{\omega}/\kappa} e^{-i q x} \right|^2 \; ,
    \label{eq:near-horizon-pex}
\end{equation}
where $q=C\tilde{\omega}+k \nu + \alpha m$,
with $k$, $C$ and $\alpha$ given by 
Eqs.~(\ref{eq:beta0}), (\ref{eq:constant-C}), and (\ref{eq:constant-alpha}), 
respectively.
 The emission-rate integral of Eq.~(\ref{eq:near-horizon-pex}) has a familiar 
 structure~\cite{acceler-rad-Schwarzschild, acceler-rad-Kerr, acceler-rad-Qopt-1, acceler-rad-Qopt-2},
 similar to that known in most 
 derivations of the Unruh effect~\cite{birrell-davies, Crispino-et-al_2008_Unruh-effect}, with a 
competition of two oscillatory factors $x^{-i\tilde{\omega}/\kappa}$ and $e^{-iqx}$. In all the related applications,
the factor $x^{-i\tilde{\omega}/\kappa}$ can be interpreted, in terms of the scale-invariant properties of CQM,
as a logarithmic singularity that oscillates wildly in the near-horizon region,
with scale invariance governed by 
 $\tilde{\omega}/\kappa = 2\Theta$.
 As a result, this coordinate singularity with a logarithmic phase governs the leading value of the integral.
Moreover, the oscillations of $e^{-iqx}$ average out to zero for large scales, making the upper limit $x_{f}$ 
effectively infinity, and 
the amplitude integral becomes
 $ \displaystyle \int_0^{x_f} dx\; x^{-i\sigma} e^{ -i q x  } 
  \approx  {q}^{-1}
  \sqrt{  2 \pi \sigma /  (e^{2 \pi \sigma} -1) }
    \; e^{i \delta} $, where $\delta$ is a real phase. 
 Thus, the final expression for the emission rate becomes
 \begin{equation}
    R_{e, {\boldsymbol{s}} }  =  
     \mathfrak{r} 
      \, g^2 k^2
       \left|\int_0^{\infty} \; dx\;  x^{-i\tilde{\omega}/\kappa} e^{-i{q} x} \right|^2 = \frac{2\pi  \mathfrak{r} \,  g^2 \tilde{\omega}}{\kappa\,\nu^2}\;\frac{1}{e^{2\pi\tilde{\omega}/\kappa}-1}\;,
    \label{eq:R_em} 
\end{equation}
where the approximation  $\nu\gg \tilde{\omega}$ applies under reasonable physical conditions.
Similarly, the absorption rate $R_{\mathrm{a}, {\boldsymbol{s}} }$ can be computed
from the second line in Eq.~(\ref{eq:P_CQM}), or simply via the replacements  $\tilde{\omega} \rightarrow -\tilde{\omega}$, and $ m \rightarrow -  m$, leading to
\begin{equation}
   R_{\mathrm{a}, {\boldsymbol{s}} }
     = \frac{2\pi \mathfrak{r} 
     g^2 \tilde{\omega}}{\kappa\,\nu^2}\;\frac{1}{1-e^{-2\pi\tilde{\omega}/\kappa}}
     = e^{2 \pi \tilde{\omega}/\kappa }  \,      R_{e, {\boldsymbol{s}} } 
     \label{eq:R_ab}
     \;  .
\end{equation}

The combined results of Eqs.~(\ref{eq:R_em}) and (\ref{eq:R_ab})
will be examined in the next section, vis-à-vis the question of thermal behavior.
In that context,
a number of noteworthy properties are satisfied by these rates.
First,
the functional form of Eq.~(\ref{eq:R_em}) is a Planck distribution,
governed by
the outgoing CQM waves in Eq.~(\ref{eq:CQM_modes}),
$ \Phi^{+ {\rm \scriptscriptstyle (CQM)}}_{\boldsymbol{s}}
\propto x^{ i\Theta} $.
Second, by contrast, the ingoing waves do not contribute, due to the cancellation of the logarithmic phases 
yielding $\sigma =0$.
A corollary of these properties
is that acceleration radiation with a Planckian distribution from a freely falling atom will exist for any generic Boulware-like state,
as a result of the nonzero conformal integral in the emission rate.
Finally, Eqs.~(\ref{eq:R_em}) and (\ref{eq:R_ab}) do not explicitly depend on the constants $k$, $C$, and $\alpha$,
i.e., the rates are insensitive to the initial conditions of the atoms.
Additional details of these properties are addressed in the appendices.

\subsection{BTZ probabilities: Thermal behavior}

The Planck functional form of Eq.~(\ref{eq:R_em}) suggests its thermal interpretation
 with the Hawking temperature~(\ref{eq:Hawking-temperature}).
This can be further examined both: (i) in terms of detailed balance and (ii) by a more comprehensive 
evaluation with the reduced field density matrix via a master equation~\cite{acceler-rad-Qopt-2}.

Detailed balance is characterized by the ratio of
 emission over absorption of photons, providing the transitions among all states needed to maintain thermal equilibrium.
This is essentially given by Eq.~(\ref{eq:R_ab}), 
which can be easily generalized for arbitrary transitions between states 
$\ket{n_{\boldsymbol{s}}}$ and $\ket{(n+1)_{\boldsymbol{s}}}$, 
with an extra factor $(n+1)$ in the rates, but the same net ratio
\begin{equation}
    \frac{R_{e, {\boldsymbol{s}} } }{R_{a, {\boldsymbol{s}} } } 
    = e^{-2\pi\tilde{\omega}/\kappa} = e^{-\beta \tilde{\omega}}
    \label{eq:ratio_em_ab_Boltzmann_2}
    \; .
\end{equation}
This is a thermal Boltzmann factor with the 
Hawking temperature~(\ref{eq:Hawking-temperature}), proportional to the surface gravity $\kappa$:
 \begin{equation}
T= \beta^{-1}  = \frac{\kappa}{2\pi} \equiv \beta_{H}^{-1} = T_H
\; 
\label{eq:temperature=Hawking}
\end{equation}
 in natural units.
 The ratio~(\ref{eq:ratio_em_ab_Boltzmann_2}) characterizes
 the detailed balance of a thermal distribution~\cite{AR_Scully-2003, AR_Belyanin-2006}
 in a form that has been extensively used for black hole 
 thermodynamics~\cite{Boltzmann-BH_1, Boltzmann-BH_2, BHT-review_Padmanabhan-2010}.
Moreover, the derivation via the near-horizon behavior 
traces the physical origin of this factor to the CQM modes, as in Eq.~(\ref{eq:near-horizon-pex}).

A more complete thermal characterization of the HBAR radiation field can be established
with the Scully-Lamb master equation for the field density 
matrix~\cite{Scully-Lamb-rho_1964, Scully-Lamb-rho_1965}, 
which can be written in a convenient multimode generalized form
\begin{equation}
\begin{aligned}
 \dot{\rho}_{\rm diag}(  \boldsymbol{ \left\{  \right. } n  \boldsymbol{\left. \right\}  } )  
 =
    - 
     \sum_{j}
     &
     \left\{
     R_{{\rm e},\, j}  \big[(n_j+1) \,
   {\rho}_{\rm diag} (  \boldsymbol{ \left\{  \right. } n  \boldsymbol{\left. \right\}  } )
      - n_j \,
   {\rho}_{\rm diag} (  \boldsymbol{ \left\{  \right. } n  \boldsymbol{\left. \right\}  }_{n_j -1} )
            \big] \right.
          \\
           &
            \left.
            +  
        R_{{\rm a},\, j} \big[ n_j \,
    {\rho}_{\rm diag} (  \boldsymbol{ \left\{  \right. } n  \boldsymbol{\left. \right\}  } )
      - (n_j + 1)  \,
     {\rho}_{\rm diag} (  \boldsymbol{ \left\{  \right. } n  \boldsymbol{\left. \right\}  }_{n_j +1} )
                           \big] \right\}
    \label{eq:master_equation_final_multimode}
    \; .
\end{aligned}
\end{equation}
Here, the index $j$ is shorthand for a given mode ${\boldsymbol{s}}_{j}  $,
with the single-mode quantum numbers $ {\boldsymbol{s}} $ chosen in an ordered sequence.
The diagonal elements of the density matrix are denoted by
$  
  {\rho}_{\rm diag} (  \boldsymbol{ \left\{  \right. } n  \boldsymbol{\left. \right\}  } )
   \equiv
  {\rho}_{  n_1,n_2, \ldots  ;   n_1,n_2, \ldots  }
$, with
$ \boldsymbol{ \left\{  \right. } n  \boldsymbol{\left. \right\}}    \equiv
\boldsymbol{ \left\{  \right.  } n_{1}, n_{2}, \ldots , n_{j } , \ldots \boldsymbol{\left. \right\}  }$ 
 referring to the occupation number representation, along with
$
  \boldsymbol{ \left\{  \right. } n  \boldsymbol{\left. \right\}}_{n_j + q}  
  \equiv
 \left\{  \right.  
 n_{1}, n_{2}, \ldots , n_{j }+ q , \ldots \boldsymbol{\left. \right\}  }
 $ (with $q$ an integer-number shift).
This is the diagonal form obtained by a partial trace from the full-fledged density matrix, and 
via an appropriate average for a cloud of freely falling atoms, with random injection times; see
   Refs.~\cite{acceler-rad-Qopt-2, QM-HBAR_centennial} for additional details.
In Eq.~(\ref{eq:master_equation_final_multimode}),
the steady-state density matrix 
$  {{\rho}}^{\mathrm (SS)}_{\rm diag}(  \boldsymbol{ \left\{  \right. } n  \boldsymbol{\left. \right\}  } )$
can be found by the vanishing of the time derivative.
The solution is given by the factorized expression
\begin{equation}
    {{\rho}}^{\mathrm (SS)}_{\rm diag}(  \boldsymbol{ \left\{  \right. } n  \boldsymbol{\left. \right\}  } )
= N \, \prod_{j}  \left(  \frac{R_{e, j } }{ R_{a, j} } \right)^{n_{j}} 
  = \frac{1}{Z}
   \prod_{j}  
    e^{-n_{j} \beta \tilde{\omega}_{j}}
    \; ,
    \label{eq:steady_state_multimode_combined}
\end{equation}
where $Z= N^{-1}= 
\prod_{j} Z_{j}
= \prod_{j} 
\left[ 1- e^{- \beta \tilde{\omega}_{j}}\right]^{-1}$ is the partition function. This is obtained using the
general form of detailed balance~(\ref{eq:ratio_em_ab_Boltzmann_2})
 needed for the ratio $  {R_{e, j } }/ R_{a, j} $. 
 Equation~(\ref{eq:steady_state_multimode_combined}) completely determines
 a thermal distribution at the Hawking temperature, including a specification of
 the steady-state average occupation numbers
$ \left\langle n_{j} \right\rangle \! ^{\! \! ^{\mathrm (SS)}}  = \left( e^{\beta \tilde{\omega}_{j}} -1 \right)^{-1}$.

With the thermal characterization, a comprehensive set of thermodynamic relations follows via the von Neumann entropy and related arguments. This is a generic property of HBAR in arbitrary geometries,
which we briefly summarize in the concluding section.

\section{Summary and Conclusions}
\label{sec:conclusions}

In this article, we derived an all-inclusive set of relations for the quantum fields and particle geodesics 
in a BTZ geometry, and their near-horizon limits. Most importantly, this exhaustive analysis shows that
 the lower-dimensional BTZ geometry offers no obstruction 
to the full realization of horizon-brightened acceleration radiation (HBAR),
including its thermal nature and emergence from near-horizon conformal quantum mechanics (CQM).

This consistent picture can be completed with a broader set of thermodynamic properties built around the HBAR entropy ${S}_{\mathcal P} $.
This is the radiation-field entropy 
(of ``scalar photons'') derived from the von Neumann entropy rate of change
$
\dot{S} = - \mathrm{Tr} \left[  \dot{{\rho}} \ln {{\rho}} \right] $; consequently,
the HBAR entropy flux has the universal form
\begin{equation}
    \dot{S}_{\mathcal P} 
    = \frac{1}{4G} \bigl| \dot{A}_{\mathcal P} \bigr|
     \label{eq:HBAR_final_nat-units}
     \; 
\end{equation}
(in natural units with $c=1$, $\hbar=1$, and $k_B=1$, while keeping $G$ explicit),
where $\bigl| \dot{A}_{\mathcal P} \bigr|$ is the absolute value of the
change in the event-horizon area due to the emission of acceleration radiation.
As the relevant near-horizon and HBAR physics in BTZ geometry was shown 
 in Sec.~\ref{sec:HBAR-thermodynamics} to be identical to that of its higher-dimensional counterparts,
  the entropy works in exactly the same manner.
Of course, this is true for both the Bekenstein-Hawking entropy and the HBAR entropy-area relation of Eq.~(\ref{eq:HBAR_final_nat-units}). Remarkably, these entropy expressions,
with the correct proportionality prefactor equal to $1/4$,
are uniquely determined via the correct temperature~(\ref{eq:temperature=Hawking})
 from the physics of acceleration radiation in a gravitational background. 
As a side note, for BTZ black holes, the ``area'' is the circumference $A = 2\pi r_+$, such that
$S = A/(4G) = \pi r_+/(2G)$. In the convention $8G=1$ used throughout this article, these relations become
$S = 4\pi r_+$ and $\dot{S}_{\mathcal P} = 2\bigl| \dot{A}_{\mathcal P} \bigr|$.

 Moreover, the temperature and entropy lead to a complete HBAR thermodynamic framework that is in one-to-one correspondence with black hole thermodynamics:
 \begin{equation}
\big( 
{S}_{\mathcal P} , {E}_{\mathcal P} ,  {J}_{{\mathcal P},z} , \text{HBAR field}
\bigr) 
\xlongleftrightarrow{\beta = \beta_{H}}
\bigl( 
S_{\mathrm{BH}} , M , J ,  \text{Hawking radiation}
\bigr) 
\; \; \;
\; ,
\label{eq:HBAR-BH-correspondence}
\end{equation}
where the photon field energy ${E}_{\mathcal P}$
and angular momentum 
 $ {J}_{{\mathcal P},z} $ can be used 
 in addition to any other relevant degrees of freedom consistent with no-hair theorems~\cite{frolov}.
 This correspondence displays a comprehensive set of robust connections between the radiation field and the 
 black hole itself--- in all dimensionalities and with any number of relevant charges---such that
  the Hawking temperature and properties of CQM provide a direct link. 
 Full details of this network of relations, 
now extended to BTZ black holes, can be found in Refs.~\cite{acceler-rad-Qopt-2, QM-HBAR_centennial}.

This paper also gives additional evidence of the effectiveness of near-horizon techniques
as practical tools leading to closed-form solutions for various aspects of black hole physics,
including the emission and absorption probabilities that define thermality in black hole backgrounds. 
This has been illustrated for acceleration radiation in generic dimensionalities,
including $(2+1)$-dimensions and the HBAR entropy in this article, but
its computational usefulness can be generalized to other problems. 

It should be noted that the HBAR models have been developed 
within the long-standing paradigm of thought experiments in theoretical physics. 
The ultimate goal is to gain deeper insights in the common realm of gravity and quantum physics.
Still, one key open question is the extent to which HBAR fields could also be explored in experimental setups. 
Astronomical black holes involving mirrors and cavities to simulate
Boulware vacua are not likely realizations, and cannot be expected to shed light on lower dimensionalities.
On the other hand, black-hole analog 
systems~\cite{BH-analogues_review-1, BH-analogues_review-2}
offer noteworthy possibilities, though concrete realizations with HBAR physics are still lacking. In particular,
there have been recent proposals of analog systems aimed at simulating BTZ black holes
using hydrodynamic~\cite{acoustic-BTZ_2022,Robin-acoustic-BTZ_2023}
and optical models~\cite{optical-BTZ_2023}.

In closing, the topics of acceleration radiation 
and HBAR entropy have attracted considerable attention in just a few years, confirming both their relevance 
and their robustness in a great variety of extended scenarios.
A partial list of examples 
 includes null-geodesic detectors~\cite{HBAR-null-geod_Majhi-2019},
 various classical extensions of black hole 
 geometries~\cite{HBAR-deSitter-BH_Bukhari-etal-2023, HBAR-BH-PFDM-halo_Bukhari-Wang-2024, HBAR-Bardeen-BH_Ovgun-Pantig-etal-2026},
 extensions to non-scalar quantum fields~\cite{HBAR-Proca-field_Pantig-Ovgun-etal-2026}
 and vibrating atoms~\cite{HBAR-vibrating-at_Pantig-Ovgun-etal-2026},
 quasinormal-mode spectroscopy~\cite{HBAR-spectroscopy_Ovgun-2026},
various higher-dimensional and stringy black holes~\cite{NH-BH-HBAR_Sen-2022, HBAR-branesBH_Das-2024},
 quantum-corrected 
geometries~\cite{QcorrBH-HBAR_Sen-2022, QcorrBH-HBAR_Jana-2024, QcorrBH-HBAR_Jana-2025, HBAR-GUP-corr-BH_Ovgun-Pantig-2026}, 
and
derivative coupling~\cite{HBAR-derivative-coupling_Das-2025, HBAR-derivative-coupling_Pantig-Ovgun-etal-2025}.
Motivated by these successful extensions,
it would be interesting to consider expanding the program to other aspects of lower-dimensional gravity.

\begin{acknowledgments}
C.R.O. and G.V.-M. were partially supported by the Army Research Office (ARO) under Grant No.~W911NF-23-1-0202. In addition, G.V.-M. gratefully acknowledges the Center for Mexican American and Latino/a Studies at the University of Houston for its generous support through a Lydia Mendoza Fellowship. H.E.C. acknowledges support by the University of San Francisco Faculty Development Fund. M.S. work was supported by U.S. Department of Energy (DE-SC-0023103, FWP-ERW7011, DE-SC0024882); Welch Foundation (A-1261); National Science Foundation (PHY-2013771); Air Force Office of Scientific Research (FA9550-20-1-0366).
\end{acknowledgments}

\appendix

\section{BTZ Geometry--Additional Properties}
\label{app:BTZ-geometry-additional}

\subsection{Metric relations and locally diagonal metric}

From Eqs.~(\ref{eq:metric-BTZ_exp})--(\ref{eq:Killing-vectors_corot-ang}), 
the following statements can be straightforwardly derived for the sector $(t ,\newphi)$:
  \begin{equation}
   g_{{t}{t}} =
    \boldsymbol{\xi}_{({t})} \cdot   \boldsymbol{\xi}_{({t})}
    \neq  \left( g^{{t}{t}} \right)^{-1}
 \; , \; \; \;
       g_{\tilde{t}\tilde{t}} =
    \boldsymbol{\xi}_{(\tilde{t})} \cdot   \boldsymbol{\xi}_{(\tilde{t})}
 = \frac{D}{g_{\newphi \newphi}}
 = \left( g^{\tilde{t}\tilde{t}} \right)^{-1} =   \left( g^{{t}{t}} \right)^{-1}
\; ,
\end{equation}
where $D=   g_{tt}  g_{\newphi \newphi} - \left(  g_{t \newphi} \right)^2$;
and for their relation to the radial sector,
  \begin{equation}
- g^{{t}{t}} =  - g^{\tilde{t}\tilde{t}} =  g_{{r}{r}} 
 \; , \; \; \;
  - g_{\tilde{t} \tilde{t}} = \left( g_{r r} \right)^{-1} =  g^{{r}{r}}
\; .
\end{equation}
Moreover, this shows that,
in the explicit form of Eq.~(\ref{eq:metric-BTZ_exp}),
there is a clear distinction between the time and radial parts of the covariant metric: 
$ -g_{tt} \neq \left( g_{rr} \right)^{-1} = f$.

\subsection{Structure of BTZ spacetime}

As summarized in Sec.~\ref{sec:BTZ-geometry_structure}, the structure of the BTZ spacetime
involves two sets of surfaces that can be identified as critical boundaries. This can be
understood from the properties of the Killing vectors 
$ \boldsymbol{\xi}_{({t})}$ and  $\boldsymbol{\xi}_{(\tilde{t})}$ 
in Eqs.~(\ref{eq:Killing-vectors_stationary}) and (\ref{eq:Killing-vectors_corot-time}),
and their relation to the metric~(\ref{eq:metric-BTZ_exp}).

First, starting sufficiently far away from the black hole, the Killing vector 
$  \boldsymbol{\xi}_{({t})} $
is timelike, but then turns into a spacelike vector for ingoing trajectories crossing the surface ${\mathcal S}$ 
given by $g_{tt} = 0$, i.e.,
where the norm of the Killing vector becomes null.
Thus, ${\mathcal S}$,
known as the static (stationary) limit or ergosurface, 
 is identified by the radius $ r_{e}$ that is the root of the equation
\begin{equation}
g_{{t}{t}} = \boldsymbol{\xi}_{({t})} \cdot   \boldsymbol{\xi}_{({t})}=0
 \Longrightarrow  
 \; \; 
  r_{e}  = \sqrt{M}l=(r_+^2+r_-^2)^{1/2}
\label{eq:static-r_{e}_02}
\; .
\end{equation}
   Then, a simple analysis shows that 
 timelike geodesics of static observers do not exist for $r <   r_{e}$, 
and all paths are dragged along in the direction of the black hole's rotation when $J \neq 0$.  
However,
outgoing geodesics can cross this ergosurface ${\mathcal S}$ 
 when $J \neq 0$; in short, 
 the condition $g_{{t}{t}} =0$ does not give a horizon because it fails to account for the effects
 of the black hole's rotational degree of freedom.
   
Second, another set of surfaces are the actual 
horizons ${\mathcal H}^{\pm}$
(outer and inner, for $r=r_{\pm}$) that
define the existence of a BTZ black hole.
 In the generic spherical-type coordinates of Eq.~(\ref{eq:metric-BTZ_exp}),  
 decreasing $r$ gradually~\cite{GR_Carroll-2003},
when the timelike $r$ = constant surface becomes null, it 
 defines a boundary where
 all timelike or null geodesics are ingoing and cannot go back to infinity. 
 At this boundary, from Eqs.~(\ref{eq:metric-BTZ})--(\ref{eq:local_ang-velocity}) ,
\begin{equation}
g^{\mu \nu} (\partial_{\mu}  r) (\partial_{\nu} r) = f = 0
\Longrightarrow   f(r) = 0 \; \; \text{or} \; \;  N^{\perp}(r) = 0  
\; 
\label{eq:horizons_from-g-radial}
\end{equation}
 (with the surface normal being proportional to $\partial_{\mu} r$).
 Moreover, as it is generally true in stationary spacetimes, the fact that
 ${\mathcal H}^{\pm}$ at $r=r_{\pm}$ is a null surface implies it is a horizon. (This is because the null direction is tangent
 to the light cone at that point, confining it to the interior side of the surface. This prevents 
 geodesics from crossing $\mathcal{H}^{\pm}$ in outgoing trajectories.)
Thus, the horizons are identified by the roots $r_{\pm}$ in Eq.~(\ref{eq:BH_horizons}), as discussed 
in Sec.~\ref{sec:BTZ-geometry_structure}.
Further analysis can be conducted with 
 the Killing vector $\boldsymbol{\xi}_{(\tilde{t})} $, which is timelike
when sufficiently close to the black hole, i.e.,
outside the event horizon, and becomes null 
when 
  \begin{equation}
 g_{\tilde{t}\tilde{t}}  = \boldsymbol{\xi}_{(\tilde{t})} \cdot   \boldsymbol{\xi}_{(\tilde{t})}
 = g_{tt} + 2 g_{t\newphi} \Omega_{H} + \Omega_{H}^2 g_{\phi \phi} =
 - f(r) + g_{\newphi \newphi} \left( \Omega_{H} - \varpi \right)^2 = 0
\; .
\label{eq:g_tilde-tt-null}
 \end{equation}
 In Eq.~(\ref{eq:g_tilde-tt-null}), the inner product $\boldsymbol{\xi}_{(\tilde{t})} \cdot   \boldsymbol{\xi}_{(\tilde{t})}$
  is computed from the definition~(\ref{eq:Killing-vectors_corot-time}); the solution is $f(r) =0$
  and $ \varpi = \Omega_{H}$, giving the outer horizon. 
  A similar analysis, redefining 
  $\tilde{\newphi}=\newphi - \Omega_{-} t$,
   can be conducted with the angular velocity
   $\Omega_{-} $ of the inner horizon.
 Therefore, this explains the criterion $f(r)=0$ leading 
 to Eq.~(\ref{eq:BH_horizons}) for the selection of Kerr horizons independently
 from the properties and null condition~(\ref{eq:horizons_from-g-radial})
  of the timelike $r$ = constant surfaces.
Furthermore, Eqs.~(\ref{eq:BH_horizons}) and (\ref{eq:g_tilde-tt-null}) select a surface $ \mathcal{H}^{\pm}$, 
with a null tangent direction along $ \boldsymbol{\xi}_{(\tilde{t})} $,
and an angular direction that is spacelike---this can be
deduced from the diagonal form of the metric of Eqs.~(\ref{eq:metric-BTZ})--(\ref{eq:local_ang-velocity}). 
 By definition, this makes $ \mathcal{H}^{\pm}$ a null surface
 generated by light rays. 
 This finding implies that the null direction is also normal to $ \mathcal{H}^{\pm}$, and
 the vanishing of its norm is fixed by the condition $g^{rr} \propto f(r) =0$,
 as shown above.
 As for the case of generalized Schwarzschild and Kerr black holes in higher dimensions, 
 such null surface is a horizon.
For the outer surface at  $r=r_{+}$, this is an event
horizon, where all timelike or null geodesics are ingoing and cannot go back to infinity; 
the inner surface at $r=r_{-}$ is classified as a Cauchy horizon~\cite{GR_Carroll-2003}.
 Finally, this simple set of arguments indicates that
 $ \boldsymbol{\xi}_{(\tilde{t})} $ describes a spacetime evolution in a corotating frame 
  with the same basic characteristics found in the neighborhood of a generalized Schwarzschild black hole
  in higher dimensions. 
Therefore, one can predict that the near-horizon physics will exhibit analogous behavior,
as in Sec.~\ref{sec:CQM_KGeq-geodesics}, governed by scale symmetry in the form of CQM.

\section{BTZ geodesics}
\label{app:BTZ-geodesics}

In this appendix, we offer a few additional details on the near-horizon approximation of BTZ geodesics, in a form 
that can be used to evaluate the transition integrals of Eq.~(\ref{eq:P_CQM}).
 
 \subsection{Radial geodesics}
 
For the radial geodesics~(\ref{eq:drdtau2}), the leading ingoing near-horizon
 expressions of Eqs.~(\ref{dtdxb0b1}) and (\ref{eq:tau-vs-x})
can be extended to their next orders by considering the expansions of $\check{e}$, $f(r)$, and $r$, so that 
$dr/d\tau$ includes the first order $\mathcal{O}(x)$. 
Specifically, from
Eqs.~(\ref{eq:local-corotating}) and (\ref{eq:corotating-energy}), 
\begin{equation}
\check{e}
\stackrel{(\mathcal{H})}{\sim}
\tilde{e} - (\! \varpi_{+} \!)' \ell x
\label{eq:check-to-tilde-e_nh}
\end{equation} 
 where 
 $ \varpi_{+}(r_{+})= \Omega_{H}$ and
 $ (\! \varpi_{+} \!)' 
 =  d\varpi_{+}(r)/dr \bigr|_{r_{+}}= -J/r_{+}^3$ 
 [from Eq.~(\ref{eq:local_ang-velocity_8G=1})].
 In addition, from Eq.~(\ref{eq:nh-expansions}),
 $f(r) \sim f'_{+} x$, so that the radial ingoing geodesic is
\begin{equation}
\frac{dr}{d\tau}
=
-
    \qty[\check{e}^2
    - f(r) \left( \frac{\ell^2}{r^2} + 1 \right)
    ]^{1/2}
    \stackrel{(\mathcal{H})}{\sim} 
   - \tilde{e} \left( 1- 2 \tilde{e} \beta_{1} x \right)^{1/2}
    \; ,
\label{eq:drdtau3}
\end{equation}
leading to
\begin{equation}
    \frac{d\tau}{dx}
    \stackrel{(\mathcal{H})}{\sim}
    - \qty(\beta_0+\beta_1 x)
    \; ,
    \label{dtdxb0b1_exp}
\end{equation}
which can be integrated to
\begin{equation}
    \tau \stackrel{(\mathcal{H})}{\sim}
 -\beta_0 x  
    + \mathcal{O}(x^2)
    + \text{constant} 
\; .
\label{eq:tau-vs-x_app}
    \end{equation}
    Here, the constants $\beta_0$ and $\beta_1$ are 
\begin{equation}
    \beta_0 
     \equiv k 
    =\frac{1}{\tilde{e}}
    \;, 
    \label{eq:beta0} 
    \end{equation}
    in agreement with Eq.~(\ref{eq:tau-vs-x}), and
\begin{equation}
    \begin{aligned}
    \beta_1 
    &
    = 
    \frac{1}{2 \tilde{e} } \,
    \left[
    \frac{2 (\! \varpi_{+} \!)'  \ell}{\tilde{e}}
    +
    \frac{ f'_{+}}{ \tilde{e}^2}
     \left( \frac{\ell^2}{r_{+}^2} + 1 \right) 
    \right]
         \\
     & = \frac{1}{\tilde{e}^3}
      \left[
      -\frac{ J \ell \tilde{e} }{r_+^3}
      +
     \kappa 
          \left( \frac{\ell^2}{r_{+}^2} + 1 \right) 
     \right]
     = \frac{1}{\tilde{e}^3}
     \left[
     \kappa + \frac{\ell(M\ell - eJ)}{r_+^3}
     \right]
    \; .
     \label{eq:beta1}
\end{aligned}
\end{equation}
The various resulting expressions can be easily transformed into each other 
via Eqs.~(\ref{eq:BH_ang-velocity}) and (\ref{eq:M-J_from_r-pm}), which amount to the
Smarr relation~\cite{Smarr-BTZ-chem_Mann-et-al}:
$M - \Omega_{H} J = \kappa r_{+}$.
It should be noted that the constants are written as inverse powers of $\tilde{e}$ 
multiplied by terms organized in terms of the parameters $\tilde{e}$ and $\ell$,
or ${e}$ and $\ell$: these are the conserved constants that encode the initial conditions.

\subsection{Coordinate-time geodesics}

For the coordinate-time geodesics, starting from  Eq.~(\ref{eq:dtdtau2}),
the next order in the lapse function is needed.
Using Eqs.~(\ref{eq:nh-expansions})--(\ref{eq:f-second-prime}),
\begin{equation}
f(r)  \stackrel{(\mathcal H)}{\sim}  f'_{+}  \, x  +  \frac{1}{2} f''_{+}   x^2 + \mathcal{O}(x^3) 
=
 f'_{+}  \, x  
\left[ 1 + \frac{1}{2} \frac{f''_{+}}{f'_{+}} x
+
\mathcal{O}(x^2) \right]  \; , 
\label{eq:nh-expansion-2nd}
\end{equation}
so that, with $\kappa = f'_{+}/2$ and Eq.~(\ref{eq:check-to-tilde-e_nh}),
\begin{align}
    \frac{dt}{d\tau}
      &
    \stackrel{(\mathcal{H})}{\sim}
    \frac{\check{e} }{2\kappa x} \, \left( 1 - \frac{1}{2} \frac{f''_{+}}{f'_{+}} x \right)
      \stackrel{(\mathcal{H})}{\sim}
    \frac{\tilde{e} }{2\kappa x} \, 
    \left( 1 -     \frac{(\! \varpi_{+} \!)' \ell x }{ \tilde {e} } \right)
    \left( 1 - \frac{1}{2} \frac{f''_{+}}{f'_{+}} x \right)
    \; ,
    \end{align}
    from which the time-to-radial-coordinate dependence is given by the chain rule, 
    along with the rate of Eq.~(\ref{dtdxb0b1_exp}),
    \begin{equation}
    -\frac{dt}{dx}
=  -\frac{dt}{d\tau} \frac{d\tau}{dx}
   \stackrel{(\mathcal{H})}{\sim}
   \left[
   \frac{ \tilde{e} }{2\kappa x}
   - \frac{1}{2 \kappa} 
   \left(
      (\! \varpi_{+} \!)' \ell 
      +
\frac{1}{2}      \frac{f''_{+}}{f'_{+}} \tilde{e}
   \right) 
   \right]
   \qty(\beta_0+\beta_1 x)
    \stackrel{(\mathcal{H})}{\sim}
    \frac{1}{2\kappa x}
    +
C +\mathcal{O}(x)
\; .
\label{eq:time-to-radial}
\end{equation}
In this expression, the constant $C$, using Eq.~(\ref{eq:beta1}) for $\beta_{1}$,
 is simply given by
\begin{equation}
C
=
 -
 \frac{1}{2}      \frac{f''_{+}}{ (f'_{+})^2 }
 +
 \frac{\beta_1\tilde{e} }{2\kappa }
+
 \frac{J }{2\kappa  r^3_{+} } \frac{\ell}{\tilde{e}}
 = 
  -
 \frac{1}{2}      \frac{f''_{+}}{ (f'_{+})^2 }
 +
 \frac{1}{2 \tilde{e}^2}
    \left( \frac{\ell^2}{r_{+}^2} + 1 \right) 
    \; ,
    \label{eq:constant-C}
\end{equation}
where the initial conditions are encoded in the second term of the rightmost side in terms
of $\tilde{e}$ and $\ell$, while the first term is purely geometrical via 
Eqs.~(\ref{eq:f-prime}) and (\ref{eq:f-second-prime}).
As it turns out, the constant $C$ in Eq.~(\ref{eq:constant-C}) can be also obtained more directly in terms of $x$,
completely bypassing the proper time, with the use of the radial geodesic~(\ref{eq:drdtau2}):
\begin{equation}
    -\frac{dt}{dx}
=  -\frac{
{dt}/{d\tau} }{ {dx}/{d\tau}
}
   \stackrel{(\mathcal{H})}{\sim}
\frac{
\check{e}/f(r)
}
{   
    \bigl[
    \check{e}^2
    - f(r) \left( {\ell^2}/{r^2} + 1 \right)
    \bigr]^{1/2}
    }
    \stackrel{(\mathcal{H})}{\sim}
    \frac{1}{2\kappa x}
    +
C +\mathcal{O}(x)
\; ,
 \label{eq:time-to-radial-direct}
\end{equation}
which gives the same results as Eqs.~(\ref{eq:time-to-radial})--(\ref{eq:constant-C}),
but without explicit use of $\beta_{1}$.
Equations~(\ref{eq:time-to-radial}) and (\ref{eq:time-to-radial-direct}) are integrated to give
\begin{equation}
t \stackrel{(\mathcal{H})}{\sim}
   - ({1}/{2\kappa } ) \ln x
    -
C x+\mathcal{O}(x^2)
\; ,
\label{eq:t-vs-x_app}
\end{equation}
 which is precisely the expression anticipated in Eq.~(\ref{eq:t-vs-x}).

\subsection{Azimuthal geodesics}

Finally, for the azimuthal geodesics, 
we can use the black hole corotating frame for a quick, direct evaluation of $\tilde{\newphi}$. This precisely
defines a constant $\alpha$ in Eq.~(\ref{eq:phi-vs-x}), which is to be used in the transition 
integrals of Eq.~(\ref{eq:P_CQM}).
Specifically, from the geodesic Eq.~(\ref{eq:dphidtau2}) for $\tilde{\newphi}$ combined with the 
 radial geodesic~(\ref{eq:drdtau2}):
\begin{equation}
    -\frac{d \tilde{\newphi}
    }{dx}
=  -\frac{
{d \tilde{\newphi}}/{d\tau} }{ {dx}/{d\tau}
}
   \stackrel{(\mathcal{H})}{\sim}
\frac{
\ell^2/r_{+}^2
}
{   
    \bigl[
    \check{e}^2
    - f(r) \left( {\ell^2}/{r^2} + 1 \right)
    \bigr]^{1/2}
    }
    \stackrel{(\mathcal{H})}{\sim}
  \frac{ \ell^2 }{ r_{+}^2 \tilde{e} }
  +\mathcal{O}(x)
\; ,
 \label{eq:azimuthal-to-radial-direct}
\end{equation}
which defines the constant
\begin{equation}
\alpha
= - \frac{ \ell^2 }{ r_{+}^2 \tilde{e} }
\; ,
    \label{eq:constant-alpha}
\end{equation}
such that 
\begin{equation}
    \tilde{\phi} \stackrel{(\mathcal H)}{\sim} \alpha x +\mo(x^2)\; . 
   \label{eq:phi-vs-x_app}
\end{equation}
It should be noted that one could more laboriously compute $\newphi$ first, from Eq.~(\ref{eq:dphidtau1}),
and then use Eq.~(\ref{eq:t-vs-x_app}) to derive $\tilde{\newphi} = \newphi - \Omega_{H} t$; 
this is a simple check that fully agrees with 
Eqs.~(\ref{eq:azimuthal-to-radial-direct})--(\ref{eq:constant-alpha}).

In conclusion, the near-horizon leading relations~(\ref{eq:tau-vs-x_app}),
(\ref{eq:t-vs-x_app}), and (\ref{eq:phi-vs-x_app}),
with the initial-condition constants
$k$, $C$, and $\alpha$ in 
Eqs.~(\ref{eq:beta0}), (\ref{eq:constant-C}), and (\ref{eq:constant-alpha}), 
give all the needed
expressions for the probability amplitudes.
In particular, this fixes the value of 
\begin{equation}
q=C\tilde{\omega}+k \nu + \alpha m
\end{equation}
in Eq.~(\ref{eq:near-horizon-pex}).
However, as highlighted in Sec.~\ref{sec:BTZ_conformal}, the final radiation formulas do not depend on 
 the specific values of these constants. This amounts to the physical insight 
 that the universal Planckian radiation is independent of the initial conditions of the atomic cloud.

\section{Exact solution for scalar field modes in the BTZ geometry}
\label{app:exact-solution_scalar-BTZ}

\subsection{General form of the exact solution}

The differential equation~\eqref{eomrmw} for a scalar field in the BTZ geometry is set up
in terms of the radial function $R_{m\omega}(r)$.
This equation has three regular singular points:
(i) the inner horizon $r=r_{-}$;
(ii) the outer horizon $r=r_{+}$;
and (iii) spatial infinity $r = \infty$.
Therefore,
there exists a transformation that converts it into 
 a hypergeometric differential equation~\cite{Abramowitz-Stegun-1972}:
\begin{equation}
    u(1-u)\frac{d^2}{du^2}
    f_{m\omega}(u)
    +\bigl[ c-(a+b+1)u \bigr]
    \frac{d}{du}
    f_{m\omega}(u)
    -ab
    f_{m\omega}(u)=0
    \; ,
    \label{eq:hypergeometric-DE}
\end{equation}
rewritten in terms of a function $f_{m\omega}(u)$.
The transformation is not unique due to the existence of multiple relations within the hypergeometric-function family.
The important requirement, which is a necessary and sufficient condition,
is that the singular points should be mapped to the values
 $u = 0$, $u = 1$, and $u = \infty$.

The general solutions to the hypergeometric differential equation 
are constructed out of the Gaussian 
 hypergeometric function (series) 
${}_{2}F_{1}(a,b;c;z)  \equiv F(a,b;c;z) $.
 As the equation has two linearly independent solutions: $f^{(1)}$ and $f^{(2)}$,
 these will be represented below with curly brackets $\left\{ f^{(1)},f^{(2)} \right\}$
 denoting linear combinations.
 At each of the three singular points $u=0, 1, \infty$, 
 there are usually two special solutions of the form $\xi^s$ times a holomorphic function of $\xi$, 
 where $\xi$ is the local variable vanishing at the singular points, with 
 $s$ being the Frobenius index or root of the indicial equation.
 This procedure applies to each of the singular points, thus generating six special solutions, 
 further related by connection formulas.
  
 The transformation of variables from Eq.~\eqref{eomrmw}
 can be easily achieved by either one of two procedures.
 The first procedure is constructive and can be built as 
  a two-step process by first defining a dimensionless 
 variable $v$ quadratic in $r$ [as suggested by the form of Eq.~(\ref{eq:f(r)_factor})],
 along with a differential equation with singular points at $v=v_{\pm}, \infty$:
 \begin{align}
 &
 v= \frac{r^2}{l^2}
 \; ,
     \label{eq:transf-v}
 \\
&
 R''(v)+ 
 [ \ln \Delta (v) ]' R'(v) 
 + 
 \frac{1}{4\Delta^2(v)}
 \left[
 m ( Mm-J\omega )-\tilde{\mu}^2l^2\Delta(v)-(m^2-\omega^2l^2)v
 \right]= 0
 \; ,
     \label{eq:radial-DE-transf-v}
 \end{align}  
 where $\Delta (v) = (v-v_{+}) (v-v_{-})$ and the primes stand for derivatives with respect $v$.
 As a second step, the variable $v$ is replaced by $u$ by rescaling and recentering around a singular point via
   a linear or fractional linear transformation.
  This second step can be performed in several ways, e.g., displacing
   $v$ around $v_{\mp}$ (one of of these) or considering the fractional ratio $(v-v_{\pm})/(v-v_{\mp})$; of course, the final results are equivalent in all of these approaches via the transformation properties
   of hypergeometric functions~\cite{Abramowitz-Stegun-1972}.
  As stated in Sec.~\ref{sec:nh-limit_from_exact-solution},
here we use the specific transformation~\cite{BTZ-Tmodes-Eucli_Ichinose-1995, ortiz2012no} 
defined by the independent-variable change
   \begin{equation}
    u=  \frac{v-v_{-}}{v_{+} - v_{-}} =  \frac{r^2-r^2_{-}}{r^2_{+} - r^2_{-}}
    \; ,
    \label{uveql1_app}
    \end{equation}
    which provides the same convenient ordering of singular points as the original equation.
    with the radial function transformation
          \begin{equation}
            R_{m\omega}(u)= 
        (u-1)^{\alpha_{+}} u^{\alpha_{-}} 
        f_{m\omega}(u)
          \label{eq:radial-DE-function-transf-u_app}
        \; ,
    \end{equation}
as in Eqs.~(\ref{uveql1})--(\ref{eq:radial-DE-function-transf-u}).

The constants $\alpha_{\pm}$
in Eq.~(\ref{eq:radial-DE-function-transf-u_app}) are imaginary parameters
 defined via Eqs.~(\ref{eq:alpha_pm})--(\ref{eq:ang-velocity_surface-gravity_both-H})
in terms of the relevant field frequency $\omega$ and the 
angular velocities $\Omega_{\pm}$
and surface gravity parameters $\kappa_{\pm}$ of both horizons.
In addition, the 
 hypergeometric differential equation parameters $(a,b,c)$ are given in Eq.~(\ref{abcdef})
 and further depend on the 
conformal weights or scaling dimensions $\Delta_{\pm}$
 of the dual CFT$_{2}$, as defined in Eq.~(\ref{eq:Delta_pm}).
    Clearly, the hypergeometric parameters~(\ref{abcdef}) are not independent complex numbers,
    being constrained by only $\Theta_{\pm}$ and $\Delta$; this yields specific conditions on the 
    solutions, as shown in the next section. 
    
The second procedure is more direct, by making use of the general properties of 
 Riemann's $P$ function~\cite{Abramowitz-Stegun-1972},
 i.e., a solution to an arbitrary equation with three regular singular points.
 Clearly, the original radial equation belongs to this family. By doing an expansion around each singular point, the Frobenius
 indices can be extracted: in Sec.~\ref{sec:nh-approx} this was done for $r=r_{+}$
 as the near-horizon framework, giving  $s^{(v_{+})}_{1,2}  = \pm i \Theta_{+}$ (the same for 
 variables $r$ and $v$);
 while the case around $r=r_{-}$ is similar, simply replacing the symbols related to 
 $r_{+}$ by $r_{-}$, i.e.,   $s^{(v_{-})}_{1,2}  = \pm i \Theta_{-}$;
  and the point at infinity has indices given by the conformal weights, as deduced directly
from the asymptotic behavior of the radial equation~(\ref{eomrmw}) or from the dual CFT$_{2}$. Thus,
one can write directly the solution as the Riemann $P$ function giving 
\begin{equation}
\! \! \! \!
 R(z)=\left(
 {
 \frac {z-z_1}{z-z_2}}\right)^{\alpha }
 \! 
 \left({\frac {z-z_3}{z-z_2}}\right)^{\gamma }
\! 
 F
 \! 
  \left(\alpha +\beta +\gamma ,\alpha +\beta '+\gamma ;1+\alpha -\alpha ' ;
 {\frac {(z-z_1)(z_3-z_2)}{(z-z_2)(z_3-z_1)}}
 \right)
\end{equation}
(Eqs.~15.6.11 and 15.6.12 in Ref.~\cite{Abramowitz-Stegun-1972}),
where the following replacements are made:
$z=v$,
$z_1=v_{-}$,
$z_3=v_{+}$,
and $z_2=v_{\infty} \rightarrow \infty$ (taking the limit and regularizing as needed)---and with indices
$\alpha, \alpha'=   s^{(v_{-})}_{1,2}$;
$\gamma, \gamma'=   s^{(v_{+})}_{1,2}$;
and $\beta, \beta'=   s^{(\infty)}_{1,2}$.
The result is identical to Eq.~(\ref{eq:radial-DE-function-transf-u_app}),
up to a constant prefactor that can be omitted by regularization
(of infinite value for $z_2=v_{\infty} \rightarrow \infty$).

  \subsection{Solutions adapted to boundary conditions}

We will briefly analyze next the behaviors of the two linearly independent solutions about each
of the two singular points most relevant for this article: corresponding to spatial infinity and the outer horizon.

\subsubsection{Asymptotic behavior and boundary condition around $r =\infty$}

To explore the asymptotic behavior when $r \to \infty$, i.e., for $u \to \infty$,
 the general solution of Eq.~(\ref{eq:hypergeometric-DE})
 is written as a linear combination of hypergeometric functions with arguments $1/u$. 
 This is a standard representation from Kummer's 24 solutions~\cite{Abramowitz-Stegun-1972}, given
 in Eq.~(\ref{eq:hypergeometric-solution_infty}).
The corresponding building blocks 
for $R_{m \omega}(u)$ are
\begin{align}
   & R^{(1)}_{m\omega}(u) 
    =
    (u-1)^{\alpha_{+}} u^{\alpha_{-} - a}
    F(a,a-c+1;a-b+1;u^{-1})
   \overset{(u \to \infty)}{\sim} u^{(\alpha_{+} +\alpha_{-} -a)}
   = u^{- s^{(\infty)}_{1}}
    \, ,
    \label{sol11}
    \\
  &  R^{(2)}_{m\omega}(u)
    =
    (u-1)^{\alpha_{+}} u^{\alpha_{-} - b}
    F(b,b-c+1;b-a+1;u^{-1})
    \overset{(u \to \infty)}{\sim} u^{(\alpha_{+} +\alpha_{-} -b)}
       = u^{ -s^{(\infty)}_{2}}
    \, ,
    \label{sol22}
\end{align}
where, from Eq.~\eqref{abcdef}, the Frobenius indices with respect to $u \sim r^2$ are
$s^{(\infty)}_{1,2} = \Delta_{\pm}/2$, 
so that $R \sim u^{- s^{(\infty)}_{1,2} }
\sim
r^{-2 s^{(\infty)}_{1,2} }
= r^{-\Delta_{\pm}}$, 
i.e., the power scaling is given by the conformal weights,
in agreement with the expected result from the dual CFT$_{2}$.
These indices can also be obtained straightforwardly
 from the spatial-infinity asymptotics of the original radial equation~(\ref{eomrmw}).
 Now, only the first index $\Delta_{+}$ has the correct behavior $s^{(\infty)}_{1}>0$
for the expected Dirichlet boundary condition at asymptotic spatial infinity. Thus,
the solution reduces to the first building block, with the correct asymptotics
\begin{equation}
    \lim_{u\to\infty}R^{(1)}_{m\omega}(u)\to 0
\label{bcfmrif1}
\; ;
\end{equation}
and the physical field modes are given by
Eq.~(\ref{nheq}).

\subsubsection{Asymptotic behavior and boundary condition around $r = r_{+}$}

The other most relevant singular point $u=1$ corresponds to the outer horizon:
$r=r_{+} $.
Exploring the solution around $u=1$ provides the all-important near-horizon regime.
This can be implemented by one of the hypergeometric connection formulas
(Eq.~15.3.9 in Ref.~\cite{Abramowitz-Stegun-1972}, with $z=1/u$):
\begin{align}
    &
    F(a,a-c+1;a-b+1;u^{-1}) =
    \nonumber\\
    &
    \frac{\Gamma(a-b+1)
    \Gamma(a+b-c)}{\Gamma(a)\Gamma(a-c+1)}(u-1)^{c-a-b}u^{a-c+1}F(1-b,1-a;c-a-b+1;1-u)
    \nonumber\\
    &
    +\frac{\Gamma(a-b+1)\Gamma(c-a-b)}{\Gamma(1-b)
    \Gamma(c-b)}
    u^aF(a,b;a+b-c+1;1-u)
    \label{eq:connection-formula-hyperg}
    \; .
\end{align}
The solution has the complete radial form~(\ref{sol11}) from the singularity at infinity,
with the values of $a$, $b$, and $c$ from Eq. \eqref{abcdef}; thus, its
  transformed expression around $u=1$ from the connection formula, Eq.~(\ref{eq:connection-formula-hyperg}),
 is
\begin{align}
\! \! \! \! 
    R_{m\omega}(u)
    &
     \! =   \!
    \frac{\Gamma(a-b+1)\Gamma(c-a-b)}{\Gamma(1-b)\Gamma(c-b)}
    (u-1)^{\alpha_{+}} u^{\alpha_{-}} 
    \!
    F(a,b;a+b-c+1;1-u)
    \nonumber\\
& \!  \! \! \! \! \! \!
+     \frac{\Gamma(a-b+1)\Gamma(a+b-c)}{\Gamma(a)\Gamma(a-c+1)}
 (u-1)^{-\alpha_{+}} u^{-\alpha_{-}} 
 F(1-b,1-a;c-a-b+1;1-u)
 \, ,
\end{align}
where, from the values of the hypergeometric parameters $(a,b,c)$ given in Eq.~(\ref{abcdef}), the conjugate relations:
$a = (1-b)^{*}$,
$b = (1-a)^{*}$,
$a -c+1= (c-b)^{*}$,
and $a+b-c =  (c-a-b)^{*}$ 
follow; and $a-b+1 = \Delta_{+} \in \mathbb{R}$. Thus, as both the gamma functions and hypergeometric
function satisfy the conjugation properties
$\left[ \Gamma(z)\right]^{*}=\Gamma(z^*)$  and $\left[ F(a,b;c;z) \right]^{*}=F(a^*,b^*; c^*;z^*)$,
and $\alpha_{\pm}= \pm i \Theta_{\pm}$ (with $\Theta_{\pm}$ real),
the two terms in Eq.~(\ref{eq:connection-formula-hyperg}) have
coefficients and hypergeometric functions that are complex conjugates of each other.
Therefore, factoring out the coefficient of the first term,
\begin{align}
\! \! \! \! \! \! \! \! 
    R_{m\omega}(u)
&
\propto
 (u-1)^{\alpha_{+}} u^{\alpha_{-}} 
   F(a,b;a+b-c+1;1-u)
   \; \nonumber\\        
        & \quad \quad \quad\quad
         +z_\gamma
       \bigl[ 
        (u-1)^{\alpha_{+}} u^{\alpha_{-}} 
   F(a,b;a+b-c+1;1-u)
        \bigr]^{*}
      \; ,
        \label{rnomlch12} 
\end{align}
where the coefficient 
\begin{equation}
    z_\gamma= 
    \frac{\Gamma(1-b)\Gamma(c-b)\Gamma(a+b-c)}{\Gamma(a)\Gamma(a-c+1)\Gamma(c-a-b)}
    =e^{-2i\gamma}
    \label{ccczg1}
    \; ,
\end{equation}
being the ratio of the complex conjugate numbers,
is a pure phase ($\gamma \in  \mathbb{R}$. 
This is indeed the expansion in Eq.~(\ref{eq:hypergeometric-solution_nh}).
Moreover, in the near-horizon limit  $x = r-r_{+} \rightarrow 0$, 
 the field modes  are given by a linear combination of the ingoing $(-)$ and outgoing $(+)$ CQM field modes from 
 Eq.~\eqref{eq:CQM_modes},
 with their amplitudes being complex conjugates of each other:
\begin{align}
    \phi_{m\omega}(t,r,\newphi) 
    =
    e^{-i\tilde{\omega} t}e^{i m \tilde{\newphi}}R_{m\omega}(x)
   & \stackrel{(\mathcal{H})}{\sim}
     e^{-i\tilde{\omega} t}e^{i m \tilde{\newphi}}
     \qty[
     e^{i\gamma}
     x^{i\Theta}+
     e^{-i\gamma}
      x^{-i\Theta}]
      \; ,
 \nonumber\\
 &=     
 e^{i\gamma}
  \Phi^{+ {\rm \scriptscriptstyle (CQM)}}_{m\tilde{\omega}}(t,x,\tilde{\newphi})
+
e^{-i\gamma} \Phi^{- {\rm \scriptscriptstyle (CQM)}}_{m\tilde{\omega}}(t,x,\tilde{\newphi})
\label{fmcqmlc1}
\; ,
\end{align}
where $ \alpha_{+} = i \Theta$ gives the exponents for the leading terms from
$u-1   \stackrel{(\mathcal H)}{\propto}  2x/\kappa l^2$.
This is the important result addressed in Sec.~\ref{sec:nh-limit_from_exact-solution}.

Finally, to implement a 
boundary condition requiring regularity of the field modes $R_{m\omega}$ in the near-horizon region, we introduce a cutoff distance $x$ from the outer horizon using 't Hooft's brick-wall regularization \cite{QBH_thooft-1985}. 
In the near-horizon limit~(\ref{fmcqmlc1}) of 
Eq.~\eqref{rnomlch12},
 a Dirichlet boundary condition reduces to
\begin{equation}
    \lim_{u\to1+x}R_{m\omega}(u)
    =0:\qquad 
    e^{i\gamma}x^{i\Theta}
     + e^{-i\gamma}x^{-i \Theta} 
     = 0,\label{dibc1}
\end{equation}
while a Neumann boundary condition yields
\begin{equation}
    \lim_{u\to1+x}\frac{dR_{m\omega}(u)}{du}
    =0:\qquad 
      e^{i\gamma}x^{i\Theta}
     - e^{-i\gamma}x^{-i \Theta} 
     \; .
     \label{dibc2}
\end{equation}
Equations~\eqref{dibc1} and \eqref{dibc2}, 
with the known value of the conformal parameter $\Theta= \tilde{\omega}/(2\kappa)$
from (\ref{eq:conformal_interaction}) and (\ref{eq:alpha_pm}),
give the following conditions for the corotating frequency $\tilde{\omega}=\omega-m\Omega_{H}$:
\begin{eqnarray}
    \text{Dirichlet:}&&\hspace{1cm}\frac{\tilde{\omega}}{\kappa}\ln(x)=\pi(2p+1)-2\gamma,\label{tosn1}\\
    \text{Neumann:}&&\hspace{1cm}\frac{\tilde{\omega}}{\kappa}\ln(x)=2\pi p-2\gamma,\label{tosn2}
\end{eqnarray}
where $p\in\mathbb{Z}$. 
This is a regularization of the frequencies with a brick-wall constraint that permits the construction of a Boulware state.
In the quantum-optics thought experiment,
this is a mathematical representation of the use of some kind of imaginary mirror near the event horizon.

Furthermore, as demonstrated in \cite{kuwata2008eigenvalue}, the conditions \eqref{tosn1} and \eqref{tosn2} can be evaluated numerically, confirming that $\tilde{\omega}=\omega-m\Omega_{H}>0$.
This constraint removes the superradiant modes, eliminating any obstructions to the proper
defintion of a Boulware state. 
We briefly discuss these issues in the next section.

\section{Superradiance and Boulware vacuum}
\label{nsprsc1}

Superradiance is a general phenomenon in scattering processes that consists of the enhancement of radiation, such
that scattered waves, under specific conditions, carry away more energy than the incident wave.
(For a comprehensive review on this topic, see \cite{DeWitt_QFT-curved, brito2015superradiance}.)

\subsection{Superradiance concepts}

Historically, superradiance can be traced back to Klein's paradox \cite{klein1929reflexion} in 1929,
 the analysis of radiation amplification due to the coherence of emitters \cite{dicke1954coherence} in 1954, and
 Zel’dovich's proof in the 1970s that radiation scattering off spinning surfaces 
 with angular velocity $\Omega$
 is amplified by the extraction of rotational energy, when its frequency $\omega$ satisfies the condition  
\begin{equation}
    \omega -m\Omega<0
    \label{suprrr}
    \; ,
\end{equation}  
where $m$ is the azimuthal number with respect to the rotational axis \cite{zel1971generation,zel1972amplification}. 
Various related phenomena leading to 
the extraction of rotational energy from a Kerr black hole, with angular velocity
$\Omega=\Omega_{H}$, are 
described by a condition of the generic form~(\ref{suprrr}); they include:
(i) the Penrose process \cite{penrose1971extraction} for ordinary particles entering the ergoregion
[with $\omega$ and $m$ in Eq.~(\ref{suprrr}) replaced by energy and angular momentum];
(ii) wave scattering of scalar fields~\cite{misner1972stability,super-rad_Starobinskii-1973},
later extended to electromagnetic and gravitational 
radiation~\cite{super-rad_Starobinskii-1974, press1972floating}, 
and spin 1/2 fields~\cite{super-rad_Starobinskii-1974, super-rad_Unruh-1974};
and (iii) a direct derivation~\cite{bekenstein1973extraction} of Eq.~(\ref{suprrr})
from Hawking's area theorem~\cite{BH-area_Hawking-1971, BH4_Bardeen-Carter-Hawking-1973}.

\subsection{Superradiance suppression in BTZ black hole}

In the analysis of the exact solution to BTZ spacetime,
 the physical solution involves field modes that vanish both at spatial infinity and in the near-horizon region 
[see Eqs.~\eqref{nheq} and \eqref{fmcqmlc1}]. 
 Furthermore, as demonstrated in \cite{kuwata2008eigenvalue}, 
this boundary condition enforces 
\begin{equation}
\tilde{\omega} = \omega - m\Omega_{H} > 0
\; ,
\end{equation}
which implies that the superradiance condition $\tilde{\omega} < 0$ cannot be satisfied.
 
 Two major conclusions of this analysis follow.
 First, the BTZ black hole exhibits no superradiant instability. 
 Second,
 as a consequence, there is no spontaneous emission in the quantum fields when the vacuum is defined
 adapted to the coordinates $(t,r, \phi)$; thus, a Boulware vacuum $\ket{0_B}$ for the scalar field
  in BTZ geometry can be defined in these coordinates.
This statement applies under reasonable boundary conditions at spatial infinity, including Dirichlet
boundary conditions.
 From the viewpoint of the HBAR thought experiment,
such vacuum, modeled via a system of mirrors (here, with appropriate boundary conditions),
provides the conditions for existence of horizon brightened acceleration radiation when atoms
are freely falling into the black hole.
 
 In addition to the argument based on the values of the corotating frequencies $\tilde{\omega}$,
 a more detailed analysis of the modes can be explicitly conducted.
  This is briefly outlined in the next section.

\subsection{Superradiance framework}

To analyze this phenomenon in the rotating BTZ black hole geometry, we follow the general method presented in \cite{matacz1993quantum,ottewill2000renormalized}, adapted to the BTZ spacetime. 

In this framework, the equation of motion for the radial function 
$R_{m\omega}(r)$ is usually written in terms
 of the tortoise coordinate $r_*(r)$, defined through
\begin{align}
    \frac{dr_*}{dr} &= N^{-2}(r)=\frac{1}{f(r)},\label{dr*dr1}\\
    r_*(r)&=\frac{l^2}{2(r_+^2-r_-^2)}\left[r_+\ln\left(\frac{r-r_+}{r+r_+}\right)-r_-\ln\left(\frac{r-r_-}{r+r_-}\right)\right].\label{rstart12}
\end{align}
Equation~(\ref{dr*dr1}) is the coordinate definition, of identical form to that used in a Schwarzschild metric;
and Eq.~(\ref{rstart12}) is the specific integrated form for the BTZ geometry.
This is a standard construction for a large variety of metrics, and as usual,
 the tortoise coordinate exhibits the following asymptotic behavior:
\begin{equation}
    r_* \to
\begin{cases}
  -\infty & \text{as } r \to r_+, \\
  \quad 0 & \text{as } r \to \infty.
\end{cases}
\label{eq:tortoise_asympt}
\end{equation}
It turns out that the mode framework can be analyzed either with $ r_*$ or $r$, but 
$ r_*$ offers a more familiar asymptotic scheme for Schrödinger-type scattering near the outer horizon, 
as the horizon is mapped from a finite value $r= r_{+}$ to 
$ r_* =-\infty$.
However, the peculiar AdS-like asymptotic behavior at spatial infinity brings the value of the
tortoise coordinate to $r^{*}=0$.
 In addition, $ r_*$ provides a starting point for the use of horizon-penetrating coordinates.
By contrast, it does not lead to a CQM description in standard form----this is why we have not
used it in the main text of this article.

 Equation~\eqref{eomrmw}
 can display scattering results more easily when it is recast into its 
 Liouville normal form without a derivative term of first order: this 
 is what may be described as a Schrödinger-like equation.
 The normal form of any equation of second order
  is achieved by a straightforward Liouville transformation that generates additional 
 non-derivative terms that can be interpreted as an effective potential.
  The normal form of
  Eq.~\eqref{eomrmw} is easily shown to be
\begin{equation}  
    \left[\frac{d^2}{dr^2_*} - V(r_*)\right]R_{m\omega}(r_*) = 0,
    \label{rfrnw1}  
\end{equation}
where we have made the substitution $R_{m\omega}(r)= \tilde{R}_{m\omega}(r)/\sqrt{r}$,
and relabelled $ \tilde{R}_{m\omega}(r)$ as $R_{m\omega}(r)$.
This amounts to the replacement  $R_{m\omega}(r)\to R_{m\omega}(r)/\sqrt{r}$ compared to the standard
form defined in Eq.~\eqref{fmnw1}, i.e., 
we employ the separation of variables
\begin{equation}
    \phi_{m\omega}(t,r,\newphi) = \frac{R_{m\omega}(r)}{\sqrt{r}}e^{-i\omega t}e^{im \newphi}
\; .
\end{equation}
As a matter of fact,
this redefined radial function is essentially
 the same as the reduced radial function $u(r) \equiv u_{m\omega}(r)$
of Sec.~\ref{sec:CQM_KGeq-geodesics}; in this section, use of
$R_{m\omega}(r)$ has notational advantages.
The resulting Schrödinger-like effective potential $V(r_*)$ is given by
\begin{eqnarray}
    V(r_*) &=& -\left(\omega - \frac{J}{2r^2} m\right)^2 + f(r) \left[\frac{m^2}{r^2} + \tilde{\mu}^2 + \frac{r^{1/2}}{2} \frac{d}{dr} \left(\frac{f(r)}{r^{3/2}}\right) + \frac{f(r)}{2r^2}\right]\nonumber\\
    &=&-\omega ^2-\frac{5 J^4}{64 r^6}-\frac{J^2}{8 l^2 r^2}+\frac{J^2 \tilde{\mu}^2}{4 r^2}
    +\frac{3 J^2 M}{8 r^4}+\frac{J m \omega }{r^2}+\frac{3 r^2}{4 l^4}+\frac{\tilde{\mu}^2 r^2}{l^2}+\nonumber\\
    &&-\frac{M}{2 l^2}+\frac{m^2}{l^2}-\tilde{\mu}^2 M-\frac{M^2}{4 r^2}-\frac{M m^2}{r^2}.
\end{eqnarray}
where the dependence $r = r(r_*)$ is implicitly assumed.
The asymptotic behavior of $    V(r_*)$ is
\begin{equation}
    V(r_*) \to
\begin{cases}
  -\tilde{\omega}^2, & r_*\to -\infty,\\
  -\omega^2 h^2_{m\omega}, & r_*\to 0,
\end{cases}
\end{equation}
in which the frequency-dependent coefficient is defined via
\begin{equation}
    h_{m\omega} = \sqrt{1 - (\omega^2l^2)^{-1} \left(m^2 + M/4\right)}.\label{w'wsq1}
\end{equation}
Thus, at $r_* \to -\infty$, we have $R(r_*) \sim e^{\pm i\tilde{\omega}r_*}$, while for $r_* \to 0$ we have $R(r_*) \sim e^{\pm i\omega h_{m\omega} r_*}$.
 Consequently, both ingoing and outgoing waves originate in either asymptotic region.

Scattering processes for an equation in normal form~(\ref{rfrnw1}) are defined in a familiar  
Schrödinger-like manner.
First, for the scattering process originating from asymptotic $r_* \to 0$, 
i.e., for an incident left mover,
the asymptotic solution of the field modes is 
\begin{equation}
        \phi^{in}_{m\omega}\sim e^{-i\omega t}e^{im \newphi}\frac{R^{+}_{m\omega}(r_*)}{\sqrt{r}}\sim\frac{1}{\sqrt{r}}
\begin{cases}
  B^+_{m\omega}e^{-i\tilde{\omega} (t+r_*)+im \tilde{\newphi}}\hspace*{4.4cm},\quad r_*\to -\infty,\\
  e^{-i\omega(t+h_{m\omega}r_*)+im \newphi} + A^+_{m\omega} e^{-i\omega(t-h_{m\omega} r_*)+im \newphi} \hspace*{.6cm},\quad r_*\to 0
  \; ,
\end{cases}
\end{equation}
where $A^+$ and $B^+$ represent the reflection and transmission amplitude coefficients, respectively 
(with an incident wave amplitude normalized to $1$).
These are designated as $in$-modes, and can be represented in the form
%
\begin{equation}
        \begin{tikzpicture}[>=stealth]
    \draw[-] (-4,0) -- (4,0) node[right]{};
    
    \filldraw (-3.8,0) circle (1.5pt) node[below]{$r_*\to-\infty$};
    \filldraw (3.8,0) circle (1.5pt) node[below]{$r_*\to 0$};
    
    \draw[->] (3,0.3) -- node[above]{$A^+_{m\omega} e^{-i\omega(t-h_{m\omega} r_*)+im \newphi}$} (4,0.3);
    \draw[<-] (3,1.5) -- node[above]{$e^{-i\omega(t+h_{m\omega}r_*)+im \newphi}$} (4,1.5);
    
    \draw[<-] (-4,0.9) -- node[above]{$B^+_{m\omega}e^{-i\tilde{\omega} (t+r_*)+im \tilde{\newphi}}$} (-3,0.9);
\end{tikzpicture}
\end{equation}
As $\omega>0$, both the incident and reflected waves possess positive energy. Furthermore, for $\tilde{\omega} > 0$ the transmitted wave carries positive energy, while for $\tilde{\omega} < 0$ it exhibits negative energy, corresponding to the superradiance phenomenon in which the energy of the reflected exceeds that of the incident wave.\\

Second, for the scattering process originating from $r_* \to -\infty$,  
i.e., for an incident right mover,
the asymptotic solution of the field modes 
for the case (i) $\tilde{\omega} > 0$, is
\begin{equation}
        \phi^{up}_{m\omega}\sim e^{-i\omega t}e^{im \newphi}\frac{R^{-}_{m\omega}(r_*)}{\sqrt{r}}\sim\frac{1}{\sqrt{r}}
\begin{cases}
  e^{-i\tilde{\omega}(t- r_*)+im \tilde{\newphi}} + A^-_{m\omega}e^{-i\tilde{\omega}(t+r_*)+im \tilde{\newphi}} \hspace*{.5cm},\quad r_*\to -\infty,\\
  B^-_{m\omega}e^{-i\omega(t- h_{m\omega} r_*)+im \newphi} \hspace*{2.8cm},\quad r_*\to 0.
\end{cases}\label{cas1min&}
\end{equation}
where $A^-$ and $B^-$ represent the reflection and transmission amplitude coefficients, respectively.
These are designated as $up$-modes, and can be represented in the form
\begin{equation}
        \begin{tikzpicture}[>=stealth]
    \draw[-] (-4,0) -- (4,0) node[right]{};
    
    \filldraw (-3.8,0) circle (1.5pt) node[below]{$r_*\to-\infty$};
    \filldraw (3.8,0) circle (1.5pt) node[below]{$r_*\to 0$};
    
    \draw[->] (-4,1.5) -- node[above]{$e^{-i\tilde{\omega}(t- r_*)+im \tilde{\newphi}}$} (-3,1.5);
    \draw[<-] (-4,0.3) -- node[above]{$A^-_{m\omega}e^{-i\tilde{\omega}(t+r_*)+im \tilde{\newphi}}$} (-3,0.3);
    
    \draw[->] (3,0.9) -- node[above]{$B^-_{m\omega}e^{-i\omega(t- h_{m\omega} r_*)+im \newphi}$} (4,0.9);
\end{tikzpicture}\label{a-disb-1}
\end{equation}
Since $\tilde{\omega} > 0$, both the incident and reflected waves carry positive energy. Similarly, as $\omega > 0$, the transmitted wave likewise exhibits positive energy.

On the other hand, for the case (ii) $\tilde{\omega} < 0$, if the $up$-mode in Eq.~\eqref{cas1min&} were used,
 the incident and reflected waves would possess negative energy, which is physically inconsistent. However, 
 through the transformations $\omega \to -\omega$ and $m \to -m$, we obtain $\tilde{\omega}\to-\tilde{\omega}=\abs{\tilde{\omega}}$, thereby ensuring positive energy for both incident and reflected waves. Consequently, the physically consistent $up$-mode for $\tilde{\omega}<0$ is given by
\begin{equation}
        \phi^{up}_{-m,-\omega}\sim e^{i\omega t}e^{-im \newphi}\frac{R^{-}_{-m,-\omega}(r_*)}{\sqrt{r}}\sim\frac{1}{\sqrt{r}}
\begin{cases}
  e^{-i\abs{\tilde{\omega}}(t- r_*)-im \tilde{\newphi}} + A^-_{-m,-\omega}e^{-i\abs{\tilde{\omega}}(t+r_*)-im \tilde{\newphi}} \hspace*{.2cm},\quad r_*\to -\infty,\\
  B^-_{-m,-\omega}e^{i\omega(t- h_{m\omega} r_*)-im \newphi} \hspace*{3.1cm},\quad r_*\to 0
  \; ,
\end{cases}
\end{equation}
which can be represented by
\begin{equation}
        \begin{tikzpicture}[>=stealth]
    \draw[-] (-4,0) -- (4,0) node[right]{};
    
    \filldraw (-3.8,0) circle (1.5pt) node[below]{$r_*\to-\infty$};
    \filldraw (3.8,0) circle (1.5pt) node[below]{$r_*\to 0$};
    
    \draw[->] (-4,1.5) -- node[above]{$e^{-i\abs{\tilde{\omega}}(t- r_*)-im \tilde{\newphi}}$} (-3,1.5);
    \draw[<-] (-4,0.3) -- node[above]{$A^-_{-m,-\omega}e^{-i\abs{\tilde{\omega}}(t+r_*)-im \tilde{\newphi}}$} (-3,0.3);
    
    \draw[->] (3,0.9) -- node[above]{$B^-_{-m,-\omega}e^{i\omega(t- h_{m\omega} r_*)-im \newphi}$} (4,0.9);
\end{tikzpicture}\label{a-disb-2}
\end{equation}
Since $\abs{\tilde{\omega}} = -\tilde{\omega} > 0$, both the incident and reflected waves carry positive energy. However, due to the transformation $\omega \to -\omega$ with $\omega > 0$, the transmitted wave now possesses negative energy, which is a direct consequence of the $\tilde{\omega} < 0$ condition.
This peculiar condition can be characterized in terms of the amplitude coefficients by
considering the Wronskian, 
which yields effective Schrödinger-like currents
that relate the amplitude coefficients.
For two linearly independent solutions, $R_1$ and $R_2$, the Wronskian satisfies
\begin{equation}
    W[R_1,R_2] = R_1 \frac{dR_2}{dr_*} - R_2 \frac{dR_1}{dr_*} = \text{constant}.\label{wr123} 
\end{equation}
Thus, for the incident left mover,
from the constancy of the Wronskian, 
\begin{equation}
1 - \abs{{A^+}}^2 = \frac{\tilde{\omega}}{{\omega h_{m\omega}}} \abs{{B^+}}^2. \label{a-wnqb--}
\end{equation}
[This can be derived by 
first 
considering $R_1 = R^+_{r_*\to 0}$ and $R_2 = \qty(R^+)^{*}_{r_*\to 0}$;
and
similarly, with $R_1 = R^+_{r_*\to -\infty}$ and $R_2 = \qty(R^+)^{*}_{r_*\to -\infty}$; and finally
making the corresponding Wronskians equal.]
Following the same reasoning for the right mover $R^-$, 
\begin{equation}
1 - \abs{{A^-}}^2 = \frac{{\omega h_{m\omega}}}{\tilde{\omega}} \abs{{B^-}}^2. \label{a-wnqb-}
\end{equation}
Moreover, through alternative choices we can obtain
\begin{equation}
    \omega B^-=\tilde{\omega} B^+\quad,\quad A^{+\,*}B^-=-\frac{\tilde{\omega}}{\omega}A^-B^{+\,*}.\label{acweq1}
\end{equation}
Thus, we conclude that, in both cases of Eqs. \eqref{a-wnqb--} and \eqref{a-wnqb-}, 
if the condition $\tilde{\omega} = \omega - m\Omega_{H} < 0$ is allowed, which corresponds to Eq.~\eqref{suprrr}, the amplitude of the reflected wave exceeds that of the incident wave, $\abs{{A^{\pm}}} > 1$, thereby manifesting the phenomenon of superradiance.
This problem originates from the mismatch of frequencies $\omega$ and $\tilde{\omega}$ associated 
with the Killing vectors $\boldsymbol\xi_{({t})} $ and $\boldsymbol\xi_{(\tilde{t})}$.

The analysis of the Kerr black hole is technically nontrivial, as superradiant modes do exist and cannot be suppressed~\cite{matacz1993quantum, ottewill2000renormalized, acceler-rad-Qopt-2}.
 (For the analysis of a massless charged scalar field on a static charged black hole geometry see \cite{balakumar2020quantum}.)
Various constructions have been considered in the literature,
including the so-called past and future Boulware vacua
$\ket{0^{\pm}_B}$, corresponding to the
 absence of particles in  
 specific portions of asymptotic infinity (past and future respectively).
 These issues were considered for the HBAR problem in Refs.~\cite{acceler-rad-Kerr} and
 \cite{acceler-rad-Qopt-2},
 with the conclusion that any generic Boulware-like state gives a Planck distribution
 for HBAR radiation.
 
 The analysis of the BTZ black hole 
 is further simplified by the absence of superradiant modes, as discussed in the previous section.
 With the framework outlined in this section,
 the key result is established from Eq.~\eqref{fmcqmlc1},
 which reveals that the near-horizon field modes consist of a superposition of ingoing and outgoing waves with complex-conjugate amplitudes. 
 This is a condition that guarantees 
 the vanishing of mode amplitudes at infinity~\cite{ortiz2012no}---in fact, this is how we derived the result in
 Sec.~\ref{sec:nh-limit_from_exact-solution}.
 With the superradiant framework of this section,
 this behavior amounts to setting $\abs{{A^-}} = 1$ in Eq.~\eqref{cas1min&}, which through Eq.~\eqref{a-wnqb-} yields $\abs{{B^-}} = 0$, precisely corresponding to the disappearance of modes at infinity.
This behavior thus implies 
   a condition on the modes that suppresses any amplitudes that could amount to superradiance.
We therefore conclude 
again that for the BTZ geometry with these boundary conditions, superradiance is manifestly absent. 
Thus, the Boulware vacuum $\ket{0_B}$ is uniquely defined for the scalar field in BTZ geometry.


\end{document}